\documentclass[seceqn]{elsart}

\usepackage{graphicx}

\usepackage{amssymb}
\usepackage{amsmath}
\usepackage{bm}

\renewcommand{\Re}{\mathop{\mathrm{Re}}\nolimits}
\newcommand{\tr}{\mathop{\mathrm{tr}}\nolimits}

\newcommand{\Ran}{\mathop{\mathrm{Ran}}\nolimits}
\newcommand{\ket}[1]{|{#1}\rangle}
\newcommand{\bra}[1]{\langle{#1}|}
\newcommand{\bras}[2]{{}_{#2}\hspace*{-0.2mm}\langle{#1}|}

\newcommand{\rint}{\int\displaylimits}
\newcommand{\tolimit}[2]{\xrightarrow{#1}}
\renewcommand{\d}{d}
\renewcommand{\e}{e}
\newcommand{\openone}{1}

\begin{document}
\begin{frontmatter}


\title{On the assumption of initial factorization in
the master equation for weakly coupled systems I: General framework}


\author[was1]{S. Tasaki},
\author[was2]{K. Yuasa\thanksref{aist}},
\author[ba1,ba]{P. Facchi},
\author[was2]{G. Kimura\thanksref{tohoku}},
\author[was2]{H. Nakazato},
\author[was2]{I. Ohba},
\author[ba2,ba]{S. Pascazio}

\address[was1]{Department of Applied Physics and Advanced Institute for Complex Systems, Waseda University, Tokyo 169-8555, Japan}
\address[was2]{Department of Physics, Waseda University, Tokyo 169-8555, Japan}
\address[ba1]{Dipartimento di Matematica, Universit\`a di Bari, I-70125 Bari, Italy}
\address[ba]{Istituto Nazionale di Fisica Nucleare, Sezione di Bari, I-70126 Bari, Italy}
\address[ba2]{Dipartimento di Fisica, Universit\`a di Bari, I-70126 Bari, Italy}
\thanks[aist]{Present address: Research Center for Information Security, National Institute of Advanced Industrial Science and Technology (AIST), 1-18-13 Sotokanda, Chiyoda-ku, Tokyo 101-0021, Japan; \textit{E-mail address:} kazuya.yuasa@aist.go.jp}
\thanks[tohoku]{Present address: Graduate School of Information Sciences, Tohoku University, Sendai 980-8579, Japan.}

\date{27 July 2006}

\begin{abstract}
We analyze the dynamics of a quantum mechanical system in
interaction with a reservoir when the initial state is not
factorized. In the weak-coupling (van Hove) limit, the dynamics can
be properly described in terms of a master equation, but a
consistent application of Nakajima--Zwanzig's projection method
requires that the reference (not necessarily equilibrium)
state of the reservoir be endowed with the mixing property.
\end{abstract}


\end{frontmatter}


\section{Introduction}
The reduced dynamics of a quantum system in contact with a
reservoir is generally described in terms of a master equation.
According to a widely accepted lore, the physical and mathematical
assumptions that are required in order to derive such an equation
are of two types: the reservoir is much larger than the system (in
a sense that can be made mathematically precise) and the coupling
between them is very weak. In these limits, the system has a
negligible influence on the reservoir and the global properties of
the latter remain unaffected during the evolution. In turns, this
enables one to assume that the reservoir is in an equilibrium
state (which can be properly defined by virtue of its macroscopic
features---the large number of degrees of freedom). One should
notice that a physically clear-cut distinction between a ``large''
reservoir and one of its subsystems can only be made because of
their (small) mutual coupling. In this respect, the two
above-mentioned hypotheses are not only consistent, but also
logically intertwined. Excellent introductions to this subject can
be found in Refs.\
\cite{ref:SpohnReview,ref:KuboTextbook,ref:WeissTextbook,ref:QuantumNoise}.

There is, however, another important requirement that is necessary
in the derivation of the master equation, but that is often taken
for granted: a factorized form of initial conditions is used to
define the dynamics. This is a hypothesis of initial statistical
independence that is certainly less easily justified. The objective
of this article is to investigate the evolution of the system with
\textit{correlated} initial conditions. We shall see that in such a case
Nakajima--Zwanzig's projection method
\cite{ref:KuboTextbook,ref:QuantumNoise,ref:Projection,ref:HaakeGeneralizedMasterEq}
requires an interesting refinement and a characterization of the
reference state of the reservoir.

Several authors have addressed the question of the modifications
that arise when it is not permissible to assume initially
independent system-environment
\cite{ref:GrabertReview,ref:Slippage,ref:Oppenheim,ref:Gorini1989,ref:Pechukas,ref:LindbladLendi,ref:Royer,ref:Buzek,ref:Kraus,ref:Benatti,ref:Hanggi2004,ref:SudarshanPRA2004,ref:OConnell,ref:ZoubiAnnPhys2004}.
In such a case, the features of the reduced dynamics have
interesting spin-offs in relation to other issues, such as the
(complete) positivity of the evolution
\cite{ref:Pechukas,ref:LindbladLendi,ref:Royer,ref:Buzek,ref:Kraus,ref:Benatti,ref:Hanggi2004,ref:SudarshanPRA2004}.

We will start off by introducing notation and setting up a general
framework in Sec.\ \ref{sec:general}. Nakajima--Zwanzig's projection
method is discussed in Sec.\ \ref{sec:dynamics} and our main result
on the evolution when the initial state is not factorized is given
in Sec.\ \ref{sec:theoremmain}. Some of the hypotheses that are
necessary in order to prove our results are thoroughly discussed in
Sec.\ \ref{sec:diagproj}. The main theorem is proved in Sec.\
\ref{sec:theorem}. In Sec.\ \ref{sec:Qpart}, we analyze the
consistency of the method and we conclude with some remarks in Sec.\
\ref{sec:Summary}. In this article, in order to stress the essential
features in a transparent way, we will restrict ourselves to the
main arguments without emphasis on mathematical rigor. A more
rigorous investigation will be presented elsewhere.

Since some of the issues discussed in this article might not be
familiar to every physicist, in particular because of the
mathematical background required to study nonequilibrium
statistical mechanics, we added some detailed explanations,
translating some more advanced mathematical concepts into rather
simple physical examples. Several details of the derivations, as
well as more tutorial issues, are discussed in Appendices
\ref{app:mixing}--\ref{app:Rlambdatau}. The tutorial discussions
are often adapted to the specific problems discussed in Secs.\
\ref{sec:theoremmain} and \ref{sec:diagproj}.

This is the first of a sequel of two papers. Here, the general
analysis is presented. In the following article \cite{ref:ArticleII}, hereafter referred
to as Article II, the main theorem on the factorization of the
state, as well as the hypotheses that are necessary for its
derivation, will be scrutinized in terms of two exactly solvable
models, in which an oscillator is coupled to a bosonic reservoir.

\section{Framework and Notation}
\label{sec:general}
We assume that the total system consists of a ``large'' reservoir B
and a ``small'' (sub)system S, so that the total Hilbert space
$\mathcal{H}_\mathrm{tot}$ can be expressed as the tensor product of
the Hilbert spaces of the reservoir $\mathcal{H}_\mathrm{B}$ and of
the system $\mathcal{H}_\mathrm{S}$,
\begin{equation}
\mathcal{H}_\mathrm{tot}
=\mathcal{H}_\mathrm{S}\otimes\mathcal{H}_\mathrm{B}.
\label{tothilb}
\end{equation}

The Liouvillian of the total system is written as
\begin{equation}
\mathcal{L}= \mathcal{L}_0 + \lambda \mathcal{L}_\mathrm{SB},
\label{Liou1}
\end{equation}
where $\lambda$ is the coupling constant and
\begin{equation}
\mathcal{L}_0
=\mathcal{L}_\mathrm{S}\otimes1_\mathrm{B}
+1_\mathrm{S}\otimes\mathcal{L}_\mathrm{B}
=\mathcal{L}_\mathrm{S}+\mathcal{L}_\mathrm{B}
\end{equation}
is the free Liouvillian, describing the free uncoupled evolutions of
the system ($\mathcal{L}_\mathrm{S}$) and of the reservoir
($\mathcal{L}_\mathrm{B}$). In the second equality, we omitted the
tensor product with the unit operator and, with an abuse of
notation, identified $\mathcal{L}_\mathrm{S}$ and
$\mathcal{L}_\mathrm{B}$ with their dilations
$\mathcal{L}_\mathrm{S}\otimes1_\mathrm{B}$ and
$1_\mathrm{S}\otimes\mathcal{L}_\mathrm{B}$, respectively. In the
following, we will always adopt such a convention, whenever no
confusion can arise. It follows that
\begin{equation}
[\mathcal{L}_\mathrm{B} , \mathcal{L}_\mathrm{S}]=0.
\label{Lioucomm}
\end{equation}
Let us write the resolution of the system Liouvillian
$\mathcal{L}_\mathrm{S}$ in terms of its eigenprojections
$\tilde{Q}_m$,
\begin{equation}
\mathcal{L}_\mathrm{S}=-i \sum_m \omega_m\tilde{Q}_m,\quad
\sum_m\tilde{Q}_m=1,\quad
\tilde{Q}_m\tilde{Q}_{n}=\delta_{mn}\tilde{Q}_m.
\label{eqn:eigenA9bis}
\end{equation}
This resolution will be used in the following. By making explicit
use of the Hamiltonians
\begin{equation}
H = H_0 + \lambda H_\mathrm{SB} = H_\mathrm{S}+ H_\mathrm{B}+\lambda
H_\mathrm{SB} \label{Hamiltonian1}
\end{equation}
and noticing that, by the definition of Liouvillian,
$\mathcal{L}\rho=-i[H,\rho]$, the superoperators $\tilde{Q}_m$ in
(\ref{eqn:eigenA9bis}) can be expressed in terms of the
eigenprojections $Q_i$ of the Hamiltonian $H_\mathrm{S}$ as
\begin{equation}\label{eqn:supordbis}
\tilde{Q}_m\rho
=\mathop{\sum_{i,j}}\limits_{E_i-E_j=\omega_m} Q_i\rho Q_j,\qquad
H_\mathrm{S} = \sum_iE_iQ_i.
\end{equation}
We are assuming that system S is finite,
$\dim\mathcal{H}_\mathrm{S}<\infty$, and thus has a pure point
spectrum.

Let $\rho(t)$ be the density matrix of the total system at time $t$.
As we anticipated, the initial state of the total system, $\rho_0$,
is usually taken to be the tensor product of a system initial state
$\rho_\mathrm{S}$ and a reservoir state $\rho_\mathrm{B}$,
\begin{equation}\label{eqn:rhoprod}
\rho_0=\rho_\mathrm{S}\otimes\rho_\mathrm{B}.
\end{equation}
This is an \textit{uncorrelated} initial state. The reservoir
state is assumed to be \textit{stationary} (with respect to the
reservoir free evolution $\mathcal{L}_\mathrm{B}$)
\begin{equation}\label{BathEq}
\mathcal{L}_\mathrm{B}\rho_\mathrm{B}=0
\end{equation}
and thus belongs to the $0$ eigenvalue of $\mathcal{L}_\mathrm{B}$.
In most applications, $\rho_\mathrm{B}=Z_\beta^{-1}\e^{-\beta
H_\mathrm{B}}$ is a thermal state at the inverse temperature
$\beta=(k_\mathrm{B}T)^{-1}$ with the normalization constant
$Z_\beta$. However, we have in mind more general instances, such as
nonequilibrium steady states \cite{ref:Ruelle}. The main target of
the present article is the reconsideration of the assumption of the
factorized initial condition (\ref{eqn:rhoprod}).

The system state $\rho_\mathrm{S}(t)$ at time $t$ is given by
\begin{equation}\label{eqn:sigtrace}
\rho_\mathrm{S}(t)=\tr_\mathrm{B}\rho(t),\qquad
\rho(t)=\e^{\mathcal{L} t}\rho_0,
\end{equation}
where $\tr_\mathrm{B}:
\mathcal{T}_1(\mathcal{H}_\mathrm{tot})\to\mathcal{T}_1(\mathcal{H}_\mathrm{S})$
is the partial trace over the reservoir degrees of freedom, a
linear operator from $\mathcal{T}_1(\mathcal{H}_\mathrm{tot})$,
the Banach space of trace-class operators on the total Hilbert
space $\mathcal{H}_\mathrm{tot}$, onto
$\mathcal{T}_1(\mathcal{H}_\mathrm{S})$. In general, unlike
$\rho(t)=\e^{-i Ht}\rho_0\e^{i Ht}$, $\rho_\mathrm{S}(t)$ is not
unitarily equivalent to $\rho_\mathrm{S}$ and the system undergoes
dissipation and/or decoherence.

\section{The Projection Method}
\label{sec:dynamics}
Consider the initial-value problem
\begin{equation}
\frac{\d}{\d t}\rho(t)=\mathcal{L}\rho(t),\qquad
\rho(0)=\rho_0,
\label{eqn:A1}
\end{equation}
where the initial density operator $\rho_0$ is \textit{not} assumed
here to be factorized like (\ref{eqn:rhoprod}). We are interested in
the reduced dynamics of the system, $\rho_\mathrm{S}(t)$ given by
(\ref{eqn:sigtrace}), with a correlated initial state $\rho_0$.

The starting point of Nakajima--Zwanzig's procedure is the
introduction of the projection operators \cite{ref:KuboTextbook,ref:QuantumNoise,ref:Projection,ref:HaakeGeneralizedMasterEq}
\begin{equation}
\label{eqn:gendefproj}
\mathcal{P}\rho=\tr_\mathrm{B}\{\rho\}\otimes\Omega_\mathrm{B}
=\sigma\otimes\Omega_\mathrm{B},\qquad
\mathcal{Q}=1-\mathcal{P},
\end{equation}
where $\rho\in\mathcal{T}_1(\mathcal{H}_\mathrm{tot})$,
$\sigma\in\mathcal{T}_1(\mathcal{H}_\mathrm{S})$, and
$\Omega_\mathrm{B}\in\mathcal{T}_1(\mathcal{H}_\mathrm{B})$ is a
given \textit{reference state} of the reservoir. Note that, from
the normalization condition $\tr_\mathrm{B}\Omega_\mathrm{B}=1$,
it follows that $\mathcal{P}^2=\mathcal{P}$ and
$\mathcal{Q}^2=\mathcal{Q}$. Therefore, $\mathcal{P}$ is the
projection onto the space of operators of the form
$\sigma\otimes\Omega_\mathrm{B}$, a subspace of
$\mathcal{T}_1(\mathcal{H}_\mathrm{tot})$ isometrically isomorphic
to $\mathcal{T}_1(\mathcal{H}_\mathrm{S})$. In particular,
\begin{equation}
\label{eqn:defproj}
\mathcal{P}\rho(t)
=\rho_\mathrm{S}(t)\otimes\Omega_\mathrm{B},\qquad
\mathcal{Q}\rho(t)=\rho(t)-\rho_\mathrm{S}(t)\otimes\Omega_\mathrm{B},
\end{equation}
where we used the definition (\ref{eqn:sigtrace}).

Since we are interested in a \textit{correlated} initial state
$\rho_0$, which is not factorized like (\ref{eqn:rhoprod}), a
question arises as to which state should be taken as the reference
state $\Omega_\mathrm{B}$
\cite{ref:HaakeGeneralizedMasterEq,ref:Oppenheim,ref:Gorini1989}
and whether the naive choice
$\Omega_\mathrm{B}=\tr_\mathrm{S}\rho_0$ is in principle
appropriate. We shall see that the situation is much more subtle
than one might naively expect: the reference state
$\Omega_\mathrm{B}$ and the reservoir dynamics cannot be
independent, but must satisfy some important properties in order
to yield a consistent description of the physical dynamics.
One of the main subjects of this article will be the specification
of the correct state $\Omega_\mathrm{B}$. Furthermore, our analysis
will show that attention should be paid to the reference state
$\Omega_\mathrm{B}$ \textit{even for a factorized initial state}.
This corroborates and sheds additional light on the rigorous
conditions for the derivation of the master equation
\cite{ref:JaksicPillet}.
At this moment, we take for granted only the stationarity
(\ref{BathEq}) of $\Omega_\mathrm{B}$ with respect to the reservoir
free dynamics, namely $\mathcal{L}_\mathrm{B}
\Omega_\mathrm{B}=0$.

Let us project the Liouville equation (\ref{eqn:A1}) onto the two
subspaces defined by $\mathcal{P}$ and $\mathcal{Q}$. Notice first
that
\begin{equation}
[\mathcal{P},\mathcal{L}_\mathrm{S}]
=[\mathcal{Q},\mathcal{L}_\mathrm{S}]=0,\quad
\mathcal{P}\mathcal{L}_\mathrm{B}
=\mathcal{L}_\mathrm{B} \mathcal{P}=0,\quad
[\mathcal{Q},\mathcal{L}_\mathrm{B}]=0.
\label{eqn:propPQ}
\end{equation}
The first equation is a consequence of the fact that
$\mathcal{L}_\mathrm{S}$ and $\mathcal{P}$ essentially operate in
different spaces, while the second derives from (\ref{BathEq}) and
from the characteristic structure of the Liouvillians,
$\tr\{\mathcal{L}\rho(t)\}=0$  (a direct consequence of probability
conservation). In addition, we require that
\begin{equation}\label{eqn:propPQ3}
\mathcal{P}\mathcal{L}_\mathrm{SB}\mathcal{P}=0,
\end{equation}
which is always satisfied as long as $H_\mathrm{SB}$ has vanishing
diagonal elements with respect to the reservoir degrees of
freedom.

By making use of (\ref{eqn:propPQ}) and (\ref{eqn:propPQ3}), the
total Liouvillian can be decomposed as
\begin{equation}
\mathcal{L}
=\mathcal{P}\mathcal{L}_\mathrm{S}\mathcal{P}
+\mathcal{Q}\mathcal{L}_0\mathcal{Q} +\lambda
\mathcal{Q}\mathcal{L}_\mathrm{SB}\mathcal{Q} +\lambda
\mathcal{P}\mathcal{L}_\mathrm{SB}\mathcal{Q} +\lambda
\mathcal{Q}\mathcal{L}_\mathrm{SB}\mathcal{P}.
\label{eqn:Liouvdec}
\end{equation}
Therefore, the free evolutions $\mathcal{L}_\mathrm{S}$ and
$\mathcal{L}_\mathrm{B}$ leave invariant the two subspaces
$\Ran\mathcal{P}$ and $\Ran\mathcal{Q}$, and all
transitions are driven by the interaction $\mathcal{L}_\mathrm{SB}$.

\section{The Main Theorem}
\label{sec:theoremmain}
The main result of this article is the following theorem, that will
be proved in Sec.\ \ref{sec:theorem}: for a correlated initial state
$\rho_0$, van Hove's ``$\lambda^2 t$'' limit
\cite{ref:SpohnReview,ref:VanHoveLimit}
of the $\mathcal{P}$-projected density operator in the
system-interaction picture,
\begin{equation}\label{eqn:densdef}
\rho_{\mathrm{I}}(\tau) = \lim_{\lambda \to 0}
\rho_\mathrm{I}^{(\lambda)}(\tau) =
 \lim_{\lambda \to 0} \e^{-\mathcal{L}_\mathrm{S}\tau/\lambda^2}
\mathcal{P} \rho(\tau/\lambda^2),
\end{equation}
is the solution of
\begin{equation}\label{eqn:inteq}
\rho_\mathrm{I}(\tau)=\mathcal{P}\rho_0
+\rint_0^\tau \d\tau'\,\mathcal{K}\rho_\mathrm{I}(\tau')
\end{equation}
with
\begin{eqnarray}
\mathcal{K}&=&\sum_m\rint_0^\infty \d t\,
\mathcal{P}\tilde{Q}_m\mathcal{L}_\mathrm{SB}\e^{\mathcal{L}_0 t}
\mathcal{L}_\mathrm{SB}\e^{-\mathcal{L}_0t}\tilde{Q}_m\mathcal{P}
\nonumber\\
&=&\sum_m\rint_0^\infty \d t\,\mathcal{P}
\tilde{Q}_m\mathcal{L}_\mathrm{SB}\e^{(\mathcal{L}_0+i\omega_m)t}
\mathcal{L}_\mathrm{SB}\tilde{Q}_m\mathcal{P}\nonumber\\
&=&-\sum_m\mathcal{P}\tilde{Q}_m\mathcal{L}_\mathrm{SB}
\frac{\mathcal{Q}}{\mathcal{L}_0+i\omega_m-0^+}
\mathcal{L}_\mathrm{SB}\tilde{Q}_m\mathcal{P},
\label{eqn:Kdef}
\end{eqnarray}
or equivalently,
\begin{equation}\label{eqn:diffeq}
\frac{\d}{\d\tau}\rho_\mathrm{I}(\tau)
=\mathcal{K}\rho_\mathrm{I}(\tau),\qquad
\rho_\mathrm{I}(0) =\mathcal{P}\rho_0
=\tr_\mathrm{B}\{\rho_0\}\otimes\Omega_\mathrm{B}.
\end{equation}
That is, even if the initial state $\rho_0$ is not in a factorized
form, but rather there is entanglement, or simply a classical
correlation, between system S and reservoir B, all correlations
disappear in van Hove's limit and system S behaves as if the total
system started from the factorized initial state in
(\ref{eqn:diffeq}) with a reservoir state $\Omega_\mathrm{B}$
specified below.

Moreover, one shows that
\begin{equation}
\lim_{\lambda\to0}\mathcal{Q}\rho(\tau/\lambda^2)
=\lim_{\lambda\to0}\left[\rho(\tau/\lambda^2)
-\tr_\mathrm{B}\{\rho(\tau/\lambda^2)\}\otimes\Omega_\mathrm{B}\right]=0,
\label{eqn:Theorem2}
\end{equation}
which makes the dynamics consistent, for no spurious term develops
in the master equation and no correlations can appear at later
times: not only the initial state, but also the state at any moment
$t$ is factorized in van Hove's limit. This supports the validity of
the assumption of the factorized state, that is frequently applied
in literature in order to derive a master equation
\cite{ref:KuboTextbook,ref:WeissTextbook,ref:QuantumNoise}. The
state of system S evolves according to the master equation
(\ref{eqn:diffeq}), while reservoir B remains in the state
$\Omega_\mathrm{B}$.

These statements are proved under the following assumptions:
\begin{enumerate}
\renewcommand{\labelenumi}{(\roman{enumi})}
\item \label{spectrum} $0$ is a simple eigenvalue of the reservoir
Liouvillian $\mathcal{L}_\mathrm{B}$ corresponding to the
eigenvector $\Omega_\mathrm{B}$ and the remaining part of the
spectrum of $\mathcal{L}_\mathrm{B}$ is absolutely continuous
\cite{ref:JaksicPillet,ref:Frohlich} (strictly speaking, for infinitely extended
reservoir, the spectrum of $\mathcal{L}_\mathrm{B}$ can be properly
defined only once the sector has been specified: in our case, the
relevant sector is that containing the state $\Omega_\mathrm{B}$);
\item the initial (correlated) state of the total system is given
 in the form
\begin{equation}
\rho_0=\Lambda(\openone_\mathrm{S}\otimes\Omega_\mathrm{B})
=\sum_iL_i(\openone_\mathrm{S}\otimes\Omega_\mathrm{B})L_i^\dag,
\label{eqn:CondInitialState}
\end{equation}
where $\Lambda$ is a bounded superoperator (i.e., $L_i$'s are
bounded operators) satisfying the normalization condition
$\tr\rho_0=1$. In other words, the initial state $\rho_0$ is a
bounded perturbation of the state
$\openone_\mathrm{S}\otimes\Omega_\mathrm{B}$ (and belongs to the
sector specified by it).
\end{enumerate}

Several comments are in order. First, observe that the Liouvillian
of an infinitely extended system can bear a point spectrum, as in
hypothesis (i) (see for instance Proposition 4.3.36 of Ref.\
\cite{BR}).

\begin{figure}[t]
\begin{center}
\includegraphics{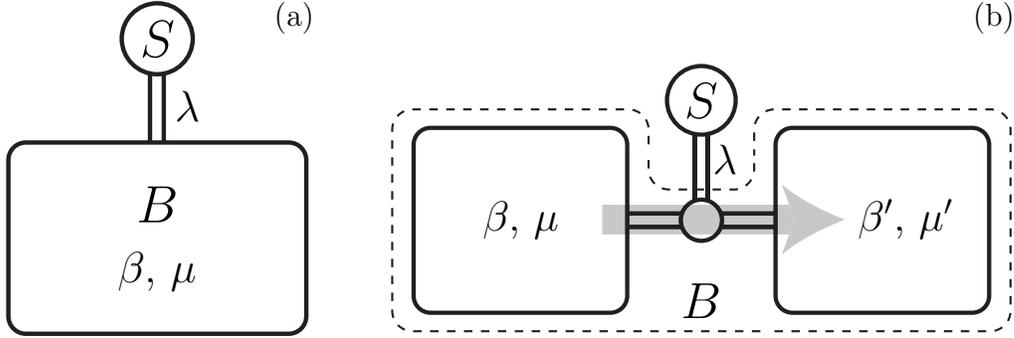}
\end{center}
\caption{Two typical examples in which the state of the reservoir is
mixing: (a) system S in contact with a reservoir B in thermal
equilibrium at the inverse temperature $\beta$ and the chemical
potential $\mu$; (b) system S in interaction with a reservoir B made
up of two subreservoirs at different temperatures $\beta$ and
$\beta'$ and chemical potentials $\mu$ and $\mu'$, with a steady
current flowing between them (NESS). In both cases, $\lambda$ is the
coupling constant: a proper application of van Hove's ``$\lambda^2
t$'' rescaling in the derivation of the master equation for S
requires that the reference state of reservoir B be mixing. }
\label{fig:ness}
\end{figure}

Second, the spectral properties required in hypothesis (i) imply
that $\Omega_\mathrm{B}$ is \textit{mixing} with respect to the
reservoir dynamics $\e^{\mathcal{L}_\mathrm{B}t}$
\cite{ref:JaksicPillet,ref:Frohlich,BR,ref:Haag,ref:ReedSimon}, i.e.,
\begin{equation}
\langle X(t)Y \rangle_{\Omega_\mathrm{B}}= \langle X
\e^{\mathcal{L}_\mathrm{B}t} Y\rangle_{\Omega_\mathrm{B}}
\to\langle X\rangle_{\Omega_\mathrm{B}}\langle
Y\rangle_{\Omega_\mathrm{B}} \quad\mathrm{as}\quad t\to\infty
\label{eqn:MixingB}
\end{equation}
for any bounded (super)operators $X$ and $Y$ of the reservoir,
where $X(t)=\e^{-\mathcal{L}_\mathrm{B}t} X
\e^{\mathcal{L}_\mathrm{B}t}$ and $\langle
X\rangle_{\rho}=\tr_\mathrm{B}\{X \rho\}$. A typical and familiar
example is the thermal equilibrium state of a free boson system at
a finite temperature, as explicitly recalled in Appendix
\ref{app:mixing}\@. Among other interesting mixing states, there
are nonequilibrium steady states (NESS), where system B consists
of several reservoirs at different temperatures and a steady
current flows among them \cite{ref:Ruelle}. These two cases are
pictorially shown in Fig.\ \ref{fig:ness}\@. Both situations are
within the scope of our analysis.

Third, hypothesis (i) can be relaxed. In fact, the proof will make
use only of the mixing property (\ref{eqn:MixingB}). Therefore, the
remaining part of the spectrum can contain a singular continuous
part: one must simply make sure that the continuous spectrum is
transient (a property automatically verified by the absolutely
continuous part, due to Riemann--Lebesgue's lemma).

Fourth, the correlation between system S and reservoir B is more
transparent if the initial state $\rho_0$ in
(\ref{eqn:CondInitialState}) is written as
\begin{equation}
\rho_0=\rho_\mathrm{S}\otimes\rho_\mathrm{B}+\delta\rho_0,
\label{eqn:CanonicalFormState}
\end{equation}
where
\begin{equation}
\rho_\mathrm{S}=\tr_\mathrm{B}\rho_0,\qquad
\rho_\mathrm{B}=\tr_\mathrm{S}\rho_0.
\end{equation}
The last term $\delta\rho_0$ represents the correlation between
system S and reservoir B\@. The important point is that each term is
a bounded perturbation of the state
$\openone_\mathrm{S}\otimes\Omega_\mathrm{B}$ and thus belongs to
the single sector specified by it (see Appendix
\ref{app:InequivSec} for a clarification and a tutorial discussion
of this issue). Indeed, the factorized part of the initial density
matrix can be written as
\begin{subequations}
\begin{align}
\rho_\mathrm{S}\otimes\rho_\mathrm{B}
&=\Lambda'(\openone_\mathrm{S}\otimes\Omega_\mathrm{B})
=\sum_{i,m,n} L_{imn}(\openone_\mathrm{S}\otimes\Omega_\mathrm{B})
L_{imn}^{\dag},
 \\
L_{imn}&=\sqrt{\rho_\mathrm{S}}
\otimes\bras{\phi_m}{\mathrm{S}}L_i\ket{\phi_n}_\mathrm{S},
\\
\intertext{and}
\delta\rho_0&=\delta\Lambda(\openone_\mathrm{S}\otimes\Omega_\mathrm{B}),\qquad
\delta\Lambda=\Lambda-\Lambda',
\label{eqn:DeltaRhoMixing}
\end{align}
\end{subequations}
where $\{\ket{\phi_n}_\mathrm{S}\}$ is any complete orthonormal
basis of system S\@. Clearly, $\delta\rho_0$ is self-adjoint,
but it is not necessarily positive-definite.

Fifth, the states of the type (\ref{eqn:CondInitialState}) belong to
a very general class: basically, if one deals with separable Hilbert
spaces, (\ref{eqn:CondInitialState}) is of the Kraus form and
therefore represents a completely positive map (that connects any
two density matrices). For quantum mechanical systems (with discrete spectra),
(\ref{eqn:CondInitialState}) covers all the possible states.  In
more general cases (with infinitely extended systems), since only bounded observables can be measured,
one usually deals with a sector, i.e.\ a set of states which are
``normal'' with respect to some reference state (cf.\ the arguments
of Secs.\ III.2 and III.3 of Ref.\ \cite{ref:Haag}).
In such a case, any normal states with respect to $1_\mathrm{S}\otimes
\Omega_\mathrm{B}$ can be written as
(\ref{eqn:CondInitialState}) with arbitrary precision.
(See also the arguments in Ref.\ \cite{ref:Frohlich}.)

Finally, the projection $\mathcal{P}$ in (\ref{eqn:defproj}) must
be defined in terms of $\Omega_\mathrm{B}$, which is mixing and
``contained'' in the initial state $\rho_0$ in the sense of
(\ref{eqn:CondInitialState}). This is the \textit{criterion} for a
\textit{consistent choice} of the reference state
$\Omega_\mathrm{B}$. We will see in the next section that
Nakajima--Zwanzig's projection $\mathcal{P}$ with the correct
reference state $\Omega_\mathrm{B}$ is nothing but the
eigenprojection of the Liouvillian $\mathcal{L}_\mathrm{B}$
belonging to the simple eigenvalue $0$, as suggested by
(\ref{eqn:propPQ}).

Note further that a state $\rho_0$ that refers [in the sense of
(\ref{eqn:CondInitialState})] to a coherent superposition of
two (or more) different mixing states of an infinite reservoir
cannot be a physical state, since it is the superposition of
states belonging to different inequivalent
sectors. 
(Imagine, for example, the superposition of states with different
temperatures.) Hypothesis (ii) is therefore reasonable from this
point of view and the states $\rho_0$ of the form
(\ref{eqn:CondInitialState}) cover diverse physically interesting
states, ranging from canonical equilibrium states to NESS, as far as
their $\Omega_\mathrm{B}$'s are mixing.

\section{The Diagonal Projection}\label{sec:diagproj}
Before we prove the theorem, let us observe that the projection
$\mathcal{P}$ with the correct reference state $\Omega_\mathrm{B}$
is nothing but the eigenprojection belonging to the simple
eigenvalue $0$ of the reservoir Liouvillian
$\mathcal{L}_\mathrm{B}$. To this end, we first show how the
eigenprojection of $0$ acts on a state of the total system.

Hypothesis (i) states that for any reservoir state
$\rho_\mathrm{B}$ in question the spectral resolution of
$\e^{\mathcal{L}_\mathrm{B}t}\rho_\mathrm{B}$ reads
\begin{equation}
\e^{\mathcal{L}_\mathrm{B}t}\rho_\mathrm{B} =\Pi_0\rho_\mathrm{B}
+\rint \e^{-i\nu t}\d\Pi(\nu)\rho_\mathrm{B},
\label{eq:specrestxt}
\end{equation}
where $\Pi_0$ and $\Pi(\nu)$ are the  spectral projections of
$\mathcal{L}_\mathrm{B}$ belonging to its simple eigenvalue $0$ and
to its absolutely continuous spectrum $\{\nu\}$, respectively.
In particular,
\begin{equation}
\mathcal{L}_\mathrm{B}\Pi_0=\Pi_0\mathcal{L}_\mathrm{B}=0,\qquad
\Pi_0^2=\Pi_0.
\label{eqn:Pi0B}
\end{equation}
By Riemann-Lebesgue's lemma, one gets from (\ref{eq:specrestxt})
\begin{equation}\label{eq:RL}
\e^{\mathcal{L}_\mathrm{B}t}\rho_\mathrm{B}
\tolimit{t\to\infty}{\longrightarrow}
\Pi_0\rho_\mathrm{B}=\Omega_\mathrm{B}
\end{equation}
in a weak sense. The last equality follows from the condition that
$\Omega_\mathrm{B}$ is the eigenstate belonging to the simple
eigenvalue 0.

As anticipated in the previous section, the requirement of an
absolutely continuous spectrum is not really necessary to prove
(\ref{eq:RL}). One needs only the mixing property
(\ref{eqn:MixingB}). Indeed, let us consider a state of the total
system belonging to the same sector of
$1_\mathrm{S}\otimes\Omega_\mathrm{B}$, i.e.\ a state $\rho$ of the
kind (\ref{eqn:CondInitialState}). For any operator $D=\sum_i
A_i\otimes X_i$, where $A_i$'s are operators of system S and $X_i$'s
bounded operators of reservoir B, the mixing property
(\ref{eqn:MixingB}) of $\Omega_\mathrm{B}$ yields
\begin{eqnarray}
\tr\{D\e^{\mathcal{L}_\mathrm{B}t}\rho\}
&=&\sum_i\tr_\mathrm{B}[X_i \e^{\mathcal{L}_\mathrm{B}t}
\tr_\mathrm{S}\{
A_i\Lambda(\openone_\mathrm{S}\otimes\Omega_\mathrm{B})
\}]\nonumber\\
& \tolimit{t\to\infty}{\longrightarrow}
&\sum_i\tr_\mathrm{B}\{X_i\Omega_\mathrm{B}\}\tr\{
A_i\Lambda(\openone_\mathrm{S}\otimes\Omega_\mathrm{B})
\}
\nonumber\\
&=&\tr[D(\tr_\mathrm{B}\{\rho\}\otimes\Omega_\mathrm{B})],
\label{eqn:DiagonalProjectionB}
\end{eqnarray}
and in this sense, we have
\begin{equation}
\e^{\mathcal{L}_\mathrm{B}t}\rho
\tolimit{t\to\infty}{\longrightarrow}
\tr_\mathrm{B}\{\rho\}\otimes\Omega_\mathrm{B} ,
\label{eqn:BathRelaxation}
\end{equation}
for any state $\rho$ of the form (\ref{eqn:CondInitialState}). By comparing (\ref{eq:RL})
and (\ref{eqn:BathRelaxation}), it is clear that the
eigenprojection $\Pi_0$ of the Liouvillian
$\mathcal{L}_\mathrm{B}$ belonging to the eigenvalue $0$ acts on
the total system as
\begin{equation}
(1_\mathrm{S}\otimes\Pi_0)\rho
=\tr_\mathrm{B}\{\rho\}\otimes\Omega_\mathrm{B}
=\mathcal{P}\rho.
\label{eqn:Pi0A}
\end{equation}
We have thus proved that (\ref{eqn:Pi0A}) holds even when the
spectrum $\{\nu\}$ in (\ref{eq:specrestxt}) contains a singular
continuous part, provided that the latter be transient, i.e.,
Riemann--Lebesgue's lemma hold also for it.

Furthermore, it is interesting to observe that the eigenprojection
$1_\mathrm{S}\otimes\Pi_0$ is nothing but the ``diagonal
projection,'' that extracts the diagonal part (with respect to the
reservoir degrees of freedom) of a density operator. To discuss the
diagonal components of a reservoir with a continuous spectrum,
consider a large parameter $\mathcal{V}$, corresponding to the
volume, so that the Hamiltonian $H_\mathrm{B}^{(\mathcal{V})}$
admits a discrete spectrum $\{E_\mu^{(\mathcal{V})}\}$, i.e.,
\begin{equation}
H_\mathrm{B}^{(\mathcal{V})} =\sum_\mu
E_\mu^{(\mathcal{V})}P_\mu^{(\mathcal{V})},\quad \sum_\mu
P_\mu^{(\mathcal{V})}=\openone,\quad
P_\mu^{(\mathcal{V})}P_\nu^{(\mathcal{V})}
=P_\mu^{(\mathcal{V})}\delta_{\mu\nu} ,
\end{equation}
with $E_\mu^{(\mathcal{V})}\neq E_\nu^{(\mathcal{V})}$ for
$\mu\neq\nu$. Then, it is easy to define the diagonal part (with
respect to the reservoir degrees of freedom) of a density matrix of
the total system, $\rho^{(\mathcal{V})}$,
\begin{equation}
\rho_\mathrm{D}^{(\mathcal{V})} =\sum_\mu P_\mu^{(\mathcal{V})}
\rho^{(\mathcal{V})}P_\mu^{(\mathcal{V})}.
\label{eqn:DiagProjDiscrete}
\end{equation}
Since $E_\mu^{(\mathcal{V})}$'s are discrete, one easily sees that
\begin{equation}
\lim_{T\to\infty}\frac{1}{T}\rint_0^T\d t\,
\e^{-i(E_\mu^{(\mathcal{V})}-E_{\mu'}^{(\mathcal{V})})t}
=\delta_{\mu\mu'},
\end{equation}
and thus,
\begin{eqnarray}
\rho_\mathrm{D}^{(\mathcal{V})}
&=&\sum_{\mu,\mu'}P_\mu^{(\mathcal{V})}
\rho^{(\mathcal{V})}P_{\mu'}^{(\mathcal{V})}\delta_{\mu\mu'}
\nonumber\\
&=&\lim_{T\to\infty}\frac{1}{T}\rint_0^T\d t\sum_{\mu,\mu'} \e^{-i
E_\mu^{(\mathcal{V})}t}P_\mu^{(\mathcal{V})} \rho^{(\mathcal{V})}
P_{\mu'}^{(\mathcal{V})}\e^{i E_{\mu'}^{(\mathcal{V})}t}
\nonumber\\
&=&\lim_{T\to\infty}\frac{1}{T}\rint_0^T\d t\, \e^{-i
H_\mathrm{B}^{(\mathcal{V})}t}
\rho^{(\mathcal{V})}\e^{i H_\mathrm{B}^{(\mathcal{V})}t}\nonumber\\
&=&\lim_{T\to\infty}\frac{1}{T}\rint_0^T\d t\,
\e^{\mathcal{L}_\mathrm{B}^{(\mathcal{V})}t}\rho^{(\mathcal{V})}.
\end{eqnarray}
We now define the diagonal part of $\rho$ via
\begin{equation}
\rho_\mathrm{D}
=\lim_{\mathcal{V}\to\infty}\rho_\mathrm{D}^{(\mathcal{V})}
=\lim_{T\to\infty}\frac{1}{T}\rint_0^T\d t\,
\e^{\mathcal{L}_\mathrm{B}t}\rho,
\label{eqn:DiagonalProjection}
\end{equation}
assuming that the two limits can be interchanged. Then, under the
assumption of the ergodicity of $\Omega_\mathrm{B}$ [which follows
from the mixing of $\Omega_\mathrm{B}$: see
Eq.\ (\ref{eqn:ergodicityA})], one has, for any state $\rho$ of the
type (\ref{eqn:CondInitialState}),
\begin{equation}
\rho_\mathrm{D}
=\tr_\mathrm{B}\{\rho\}\otimes\Omega_\mathrm{B},
\label{eqn:DiagonalProjectionLocal}
\end{equation}
which is the right-hand side of (\ref{eqn:Pi0A}). Indeed, for any
bounded operator $D=\sum_i A_i\otimes X_i$ considered in
(\ref{eqn:DiagonalProjectionB}),
\begin{eqnarray}
\tr\{D\rho_\mathrm{D}\}
&=&\sum_i\lim_{T\to\infty}\frac{1}{T}\rint_0^T\d t
\tr_\mathrm{B}[X_i \e^{\mathcal{L}_\mathrm{B}t}
\tr_\mathrm{S}\{A_i\Lambda(\openone_\mathrm{S}\otimes\Omega_\mathrm{B})\}]\nonumber\\
&=&\sum_i\tr_\mathrm{B}\{X_i\Omega_\mathrm{B}\}
\tr\{A_i\Lambda(\openone_\mathrm{S}\otimes\Omega_\mathrm{B})\}\nonumber\\
&=&\tr[D(\tr_\mathrm{B}\{\rho\}\otimes\Omega_\mathrm{B})],
\end{eqnarray}
which proves (\ref{eqn:DiagonalProjectionLocal}), and the
eigenprojection of $\mathcal{L}_\mathrm{B}$ belonging to the
discrete eigenvalue $0$ is the diagonal projection.
Notice that the ergodicity of $\Omega_\mathrm{B}$ is sufficient to
show that $1_\mathrm{S}\otimes\Pi_0$ is the diagonal projection. A
direct demonstration of the diagonal projection
(\ref{eqn:DiagonalProjectionLocal}) in a simple model is given in
Appendix \ref{app:DiagonalProjection}\@.

It is now clear that the eigenprojection $1_\mathrm{S}\otimes\Pi_0$
in (\ref{eqn:Pi0A}) acts like the projection $\mathcal{P}$ defined
in (\ref{eqn:defproj}), provided the state $\rho(t)$ is of the form
$\Lambda_t(\openone_\mathrm{S}\otimes\Omega_\mathrm{B})$, with a
bounded superoperator $\Lambda_t$. Since $\rho(t)$ has evolved from
the initial state (\ref{eqn:CondInitialState}), it always satisfies
this criterion. As we will discuss in the following, the
eigenprojection $1_\mathrm{S}\otimes\Pi_0$ enables us to deal with
the point spectrum and yields the right choice for the projection
$\mathcal{P}$ in order to derive a master equation in van Hove's
limit.

Summarizing, the initial state $\rho_0$ ``contains,'' in the sense
of (\ref{eqn:CondInitialState}) and
(\ref{eqn:BathRelaxation}), the mixing state $\Omega_\mathrm{B}$.
The theorem stated in Sec.\ \ref{sec:theoremmain} will be proved
in Sec.\ \ref{sec:theorem} with the projection operator
\begin{equation}
\mathcal{P}=1_\mathrm{S}\otimes\Pi_0,
\end{equation}
which is the eigenprojection of
$1_\mathrm{S}\otimes\mathcal{L}_\mathrm{B}$ belonging to
the unique simple eigenvalue $0$. In the following, with an abuse
of notation, $1_\mathrm{S}\otimes\Pi_0$ will be written simply as
$\Pi_0$, as far as no confusion can arise.

Some additional comments are in order. If we assume only the
ergodicity of $\Omega_\mathrm{B}$, rather than mixing, there can be
other eigenvalues different from $0$. 
The point spectrum
other than $0$ is out of control and, as we shall see in the
following section, provokes the appearance of a secular term.

It is worth noting that the mixing property of
$\Omega_\mathrm{B}$ is crucial even for a
\textit{factorized} initial state like (\ref{eqn:rhoprod}),
although this point is usually not thoroughly discussed. (This is
an interesting byproduct of our analysis, that is rather motivated
by the study of nonfactorized initial states.) As we shall see, a
wrong projection $\mathcal{P}$ would give rise to a divergence
that has nothing to do with the initial correlation.

In the
standard derivations of the master equation for a factorized
initial state, the projection (\ref{eqn:defproj}) is defined in
terms of the same $\Omega_\mathrm{B}$ which is contained in the
factorized initial state (\ref{eqn:rhoprod}) and is usually the
canonical state at temperature $T$, that is a mixing state. The
criteria listed in Sec.\ \ref{sec:theoremmain} are thus satisfied
and the projection is properly chosen to be
$\mathcal{P}=1_\mathrm{S}\otimes\Pi_0$, provoking no problem. In
this standard situation, the choice of the reference state is
obvious: probably this often induces one to assume that
Nakajima-Zwanzig's reference state is simply the canonical one
(and therefore need not be characterized). However, in more
articulated situations, such as those of NESS (see Fig.\
\ref{fig:ness}), the reference state must be chosen with care: our
analysis shows that a correct reference state must in general be
mixing.

\section{Proof of the Theorem}
\label{sec:theorem}
Let us now prove the theorem stated in Sec.\ \ref{sec:theoremmain}\@.
By projecting the Liouville equation (\ref{eqn:A1}) onto the two
subspaces defined by $\mathcal{P}$ and $\mathcal{Q}$, and using
(\ref{eqn:Liouvdec}), one gets
\begin{subequations}
\begin{eqnarray}\label{eqn:A3}
\frac{\d}{\d t}\mathcal{P}\rho &=\mathcal{L}_\mathrm{S}
\mathcal{P}\rho
+\lambda\mathcal{P}\mathcal{L}_\mathrm{SB}\mathcal{Q}\rho,
\label{eqn:A3A}\\
\frac{\d}{\d t}\mathcal{Q}\rho &=\mathcal{L}'_0\mathcal{Q}\rho
+\lambda\mathcal{Q}\mathcal{L}_\mathrm{SB}\mathcal{P}\rho,
\label{eqn:A3B}
\end{eqnarray}
\end{subequations}
respectively, where
\begin{equation}\label{eqn:cLp0}
\mathcal{L}'_0
=\mathcal{L}_0+\lambda{\mathcal
Q}\mathcal{L}_\mathrm{SB}\mathcal{Q}.
\end{equation}
By formally integrating out the second equation and plugging the
result into the first one, one gets the following \textit{exact}
equation for the $\mathcal{P}$-projected operator in the interaction
picture \cite{ref:HaakeGeneralizedMasterEq}
\begin{equation}
\frac{\d}{\d t}\e^{-\mathcal{L}_\mathrm{S}t}\mathcal{P}\rho(t)
=\lambda^2\rint_0^t\d t'\,\e^{-\mathcal{L}_\mathrm{S}t}
\mathcal{P}\mathcal{L}_\mathrm{SB}\e^{\mathcal{L}_0'(t-t')}
\mathcal{L}_\mathrm{SB}\mathcal{P}\rho(t')
+\lambda \e^{-\mathcal{L}_\mathrm{S}t}
\mathcal{P}\mathcal{L}_\mathrm{SB}\e^{\mathcal{L}'_0t}
\mathcal{Q}\rho_0.
\label{eqn:A4}
\end{equation}
The last term in the right-hand side represents the contribution
arising from a possible initial correlation between system S and
reservoir B\@. We will show that this term dies out in the
weak-coupling limit $\lambda\to0$ with the scaled time
$\tau=\lambda^2t\,(>0)$ fixed,
\textit{provided the projection $\mathcal{P}$ is chosen to be the
eigenprojection $1_\mathrm{S}\otimes\Pi_0$ belonging to the simple
eigenvalue $0$ of $\mathcal{L}_\mathrm{B}$}. To this end, consider
the density operator
\begin{equation}\label{eq:rhoIlambda}
\rho_\mathrm{I}^{(\lambda)}(\tau) =
\e^{-\mathcal{L}_\mathrm{S}\tau/\lambda^2}\mathcal{P}
\rho(\tau/\lambda^2),
\end{equation}
introduced in (\ref{eqn:densdef}), that for any nonvanishing
$\lambda$ satisfies
\begin{eqnarray}
\frac{\d}{\d\tau}\rho_\mathrm{I}^{(\lambda)}(\tau)
&=&\rint_0^{\tau/\lambda^2}\d t\,
\e^{-\mathcal{L}_\mathrm{S}\tau/\lambda^2}\mathcal{P}
\mathcal{L}_\mathrm{SB}\e^{\mathcal{L}'_0(\tau/\lambda^2-t)}
\mathcal{L}_\mathrm{SB}\mathcal{P}\e^{\mathcal{L}_\mathrm{S}t}
\rho_\mathrm{I}^{(\lambda)}(\lambda^2t)\nonumber\\
&&{}+\frac{1}{\lambda}\e^{-\mathcal{L}_\mathrm{S}\tau/\lambda^2}
\mathcal{P}\mathcal{L}_\mathrm{SB}
\e^{\mathcal{L}'_0 \tau/\lambda^2}\mathcal{Q}\rho_0,
\label{eqn:A6}
\end{eqnarray}
with the initial condition
\begin{equation}
\rho_\mathrm{I}^{(\lambda)}(0)=\mathcal{P}\rho_0.
\label{eqn:condiniz}
\end{equation}
By integrating (\ref{eqn:A6}), one gets
\begin{eqnarray}
\rho_\mathrm{I}^{(\lambda)}(\tau)
=\mathcal{P}\rho_0
&+&\rint_0^\tau \d\tau'
\rint_0^{\tau'/\lambda^2}\d t\,
\e^{-\mathcal{L}_\mathrm{S}\tau'/\lambda^2}
\mathcal{P}\mathcal{L}_\mathrm{SB}
\e^{\mathcal{L}'_0(\tau'/\lambda^2-t)}\mathcal{L}_\mathrm{SB}
\mathcal{P}\e^{\mathcal{L}_\mathrm{S}t}
\rho_\mathrm{I}^{(\lambda)}(\lambda^2t)\nonumber\\
&+&\frac{1}{\lambda}\rint_0^\tau \d\tau'\,
\e^{-\mathcal{L}_\mathrm{S}\tau'/\lambda^2}
\mathcal{P}\mathcal{L}_\mathrm{SB}
\e^{\mathcal{L}'_0\tau'/\lambda^2}\mathcal{Q}\rho_0.
\label{eqn:integriniz}
\end{eqnarray}
The second term is rearranged as
\begin{eqnarray}
& & \rint_0^\tau \d\tau'\rint_0^{\tau'/\lambda^2}\d t\,
\e^{-\mathcal{L}_\mathrm{S}\tau'/\lambda^2}\mathcal{P}
\mathcal{L}_\mathrm{SB}\e^{\mathcal{L}_0'(\tau'/\lambda^2-t)}
\mathcal{L}_\mathrm{SB}\mathcal{P}\e^{\mathcal{L}_\mathrm{S}t}
\rho_\mathrm{I}^{(\lambda)}(\lambda^2t)\nonumber\\
& & =\rint_0^{\tau/\lambda^2}\d t\rint_0^{\lambda^2t}\d\tau'\,
\e^{-\mathcal{L}_\mathrm{S}t}\mathcal{P}\mathcal{L}_\mathrm{SB}
\e^{\mathcal{L}_0'(t-\tau'/\lambda^2)}\mathcal{L}_\mathrm{SB}
\mathcal{P}\e^{\mathcal{L}_\mathrm{S}\tau'/\lambda^2}
\rho_\mathrm{I}^{(\lambda)}(\tau')\nonumber\\
& & =\rint_0^{\tau}\d\tau'
\rint_{\tau'/\lambda^2}^{\tau/\lambda^2}\d t\,
\e^{-\mathcal{L}_\mathrm{S}t}\mathcal{P}\mathcal{L}_\mathrm{SB}
\e^{\mathcal{L}_0'(t-\tau'/\lambda^2)}\mathcal{L}_\mathrm{SB}
\mathcal{P}\e^{\mathcal{L}_\mathrm{S}\tau'/\lambda^2}
\rho_\mathrm{I}^{(\lambda)}(\tau')\nonumber\\
& & =\rint_0^{\tau}\d\tau'\rint_0^{(\tau-\tau')/\lambda^2}\d t\,
\e^{-\mathcal{L}_\mathrm{S}(t+\tau'/\lambda^2)}\mathcal{P}
\mathcal{L}_\mathrm{SB}\e^{\mathcal{L}_0't}\mathcal{L}_\mathrm{SB}
\mathcal{P}\e^{\mathcal{L}_\mathrm{S}\tau'/\lambda^2}
\rho_\mathrm{I}^{(\lambda)}(\tau')\nonumber\\
& & =\sum_{m,n}\rint_0^{\tau}\d\tau'\,
\e^{i(\omega_m-\omega_n)\tau'/\lambda^2}
\mathcal{K}_{mn}^{(\lambda)}(\tau-\tau')
\rho_\mathrm{I}^{(\lambda)}(\tau') ,
\end{eqnarray}
with the memory kernel
\begin{equation}
\mathcal{K}_{mn}^{(\lambda)}(\tau) =\rint_0^{\tau/\lambda^2}\d
t\,\mathcal{P}\tilde{Q}_m
\mathcal{L}_\mathrm{SB}\mathcal{Q}\e^{(\mathcal{L}_0'+i\omega_m)t}
\mathcal{L}_\mathrm{SB}\tilde{Q}_n\mathcal{P}.
\label{eqn:Memory}
\end{equation}
The last term in Eq.\ (\ref{eqn:integriniz}), which is relevant to
the initial correlation, reads
\begin{eqnarray}
\mathcal{I}^{(\lambda)}(\tau)
&=\frac{1}{\lambda}\rint_0^\tau \d\tau'\,
\e^{-\mathcal{L}_\mathrm{S}\tau'/\lambda^2}\mathcal{P}
\mathcal{L}_\mathrm{SB}\e^{\mathcal{L}_0'\tau'/\lambda^2}
\mathcal{Q}\rho_0\nonumber\\
&=\lambda\sum_m\rint_0^{\tau/\lambda^2}\d t\,\mathcal{P}
\tilde{Q}_m\mathcal{L}_\mathrm{SB}\e^{(\mathcal{L}_0'+i\omega_m)t}
\mathcal{Q}\rho_0.
\label{eqn:CorrelationTerm}
\end{eqnarray}
In conclusion,
\begin{equation}
\rho_\mathrm{I}^{(\lambda)}(\tau)
=\mathcal{P}\rho_0 +\sum_{m,n}\rint_0^{\tau}\d\tau'\,
\e^{i(\omega_m-\omega_n)\tau'/\lambda^2}
\mathcal{K}_{mn}^{(\lambda)}(\tau-\tau')
\rho_\mathrm{I}^{(\lambda)}(\tau')
+\mathcal{I}^{(\lambda)}(\tau).
\label{eqn:ReducedDynamics}
\end{equation}

We can now analyze the van Hove limits of the memory kernel
$\mathcal{K}_{mn}^{(\lambda)}(\tau)$ and of the initial correlation
$\mathcal{I}^{(\lambda)}(\tau)$. Both limits can be computed at the
same time if we consider the van Hove limit of the operator
\begin{equation}
\mathcal{R}_m^{(\lambda)}(\tau)
=\rint_0^{\tau/\lambda^2}\d t\,\mathcal{Q}
\e^{(\mathcal{L}_0'+i\omega_m)t},
\label{eqn:KEY}
\end{equation}
since both $\mathcal{K}_{mn}^{(\lambda)}(\tau)$ and
$\mathcal{I}^{(\lambda)}(\tau)$ contain it. The analysis of the
properties of $\mathcal{R}_m^{(\lambda)}(\tau)$ is given in Appendix
\ref{app:Rlambdatau}, where it is shown that, irrespectively of the
point spectrum of $\mathcal{L}_0'$, one obtains
\begin{equation}
\mathcal{R}_m^{(0)}=\lim_{\lambda\to0}\mathcal{R}_m^{(\lambda)}(\tau)
=-\frac{\mathcal{Q}}{\mathcal{L}_0+i\omega_m-0^+}\quad(\tau>0),
\label{eqn:KeyFormula}
\end{equation}
\textit{provided that
$\mathcal{P}=1_\mathrm{S}\otimes\Pi_0$}. The expression
(\ref{eqn:KeyFormula}) is our key formula: it immediately leads us
to the conclusion that, in van Hove's limit, the memory kernel
$\mathcal{K}_{mn}^{(\lambda)}(\tau)$ in (\ref{eqn:Memory}) is
reduced to a Markovian generator
\begin{eqnarray}
\mathcal{K}_{mn}^{(\lambda)}(\tau)
&=&\mathcal{P}\tilde{Q}_m\mathcal{L}_\mathrm{SB}
\mathcal{R}_m^{(\lambda)}(\tau)\mathcal{L}_\mathrm{SB}
\tilde{Q}_n\mathcal{P}\nonumber\\
&\tolimit{\lambda\to 0}{\longrightarrow}& {}
\mathcal{K}_{mn}^{(0)}=-\mathcal{P}\tilde{Q}_m\mathcal{L}_\mathrm{SB}
\frac{\mathcal{Q}}{\mathcal{L}_0+i\omega_m-0^+}
\mathcal{L}_\mathrm{SB}\tilde{Q}_n\mathcal{P} ,
\end{eqnarray}
while the correlation term $\mathcal{I}^{(\lambda)}(\tau)$ in
(\ref{eqn:CorrelationTerm}) disappears
\begin{eqnarray}
\mathcal{I}^{(\lambda)}(\tau)
&=\lambda\sum_m\mathcal{P}\tilde{Q}_m\mathcal{L}_\mathrm{SB}
\mathcal{R}_m^{(\lambda)}(\tau)\mathcal{Q}\rho_0
\tolimit{\lambda\to 0}{\longrightarrow}
0, 
\end{eqnarray}
so that the reduced dynamics (\ref{eqn:ReducedDynamics}) becomes
\begin{eqnarray}
\rho_\mathrm{I}^{(\lambda)}(\tau) &\tolimit{\lambda\to
0}{\longrightarrow}&  \mathcal{P}\rho_0
+\sum_{m,n}\rint_0^{\tau}\d\tau'\,\delta_{mn}
\mathcal{K}_{mn}^{(0)} \rho_\mathrm{I}(\tau')
\nonumber\\
&=&\mathcal{P}\rho_0
 + \rint_0^{\tau}\d\tau'\,\mathcal{K}\rho_\mathrm{I}(\tau').
\end{eqnarray}
The master equation in van Hove's limit therefore reads
\begin{subequations}
\begin{eqnarray}
& & \frac{\d}{\d\tau}\rho_\mathrm{I}(\tau)
=\mathcal{K}\rho_\mathrm{I}(\tau), \\ & & \mathcal{K} =\sum_m
\mathcal{K}_{mm}^{(0)}
=-\sum_m\mathcal{P}\tilde{Q}_m\mathcal{L}_\mathrm{SB}
\frac{\mathcal{Q}}{\mathcal{L}_0+i\omega_m-0^+}
\mathcal{L}_\mathrm{SB}\tilde{Q}_m\mathcal{P}.
\end{eqnarray}
\end{subequations}
We have thus proved (\ref{eqn:inteq})--(\ref{eqn:diffeq}), the
first part of the theorem in Sec.\ \ref{sec:theoremmain}. In van
Hove's limit, the density matrix evolves as if it started from the
initial condition $\mathcal{P} \rho_0$, even when
$\mathcal{Q}\rho_0\neq0$: the initial correlation dies out
immediately (at $\tau=0^+$) in the scaled time $\tau$. The
contribution originating from the initial correlation between the
system and the reservoir disappears in the scaling limit and one
is allowed to start from an initial density matrix in the
factorized form (\ref{eqn:diffeq}). From a physical point of view,
the factorization Ansatz described above simply means that the
``initial'' correlations between the system and its environment
are ``forgotten'' on a time scale of order $\lambda^2$ in $\tau$
(which is very small when compared to the timescale of the
evolution of the system). We shall see in Article II \cite{ref:ArticleII}, by looking
at a specific example, that the problem of the relevant timescales
must be tackled with care, as it also involves locality issues
related to the ``size'' of the local observables of the reservoir.

It is important to note that, if we choose a different projection
$\mathcal{P}$ from the eigenprojection of $\mathcal{L}_\mathrm{B}$
belonging to its vanishing eigenvalue, and fail to appropriately
remove the point spectrum of $\mathcal{L}_\mathrm{B}$, Eq.\
(\ref{eqn:KeyFormula}) does not hold anymore, and
$\mathcal{R}_m^{(\lambda)}(\tau)$ diverges in van Hove's limit.
[See, for example, Eq.\ (\ref{eqn:ContPointSpec}) in Appendix
\ref{app:Rlambdatau}.] $\mathcal{K}_{mn}^{(\lambda)}(\tau)$ and
$\mathcal{I}^{(\lambda)}(\tau)$ accordingly diverge and the van Hove
limit of the master equation does not exist. This is because a
``wrong'' projection $\mathcal{P}$ would project the reservoir onto
a wrong (in general non-stationary) state, so that the system
evolution would develop a secular term in $\tau/\lambda^2$ [like in
Eq.\ (\ref{eqn:ContPointSpec})]. It is remarkable that such a
secular term appears as a consequence of a sloppy application of
Nakajima--Zwanzig's projection method: in a sense, the very method
makes sense only if applied to the ``correct'' mixing state.

As stressed at the end of the previous section, the reference state
of the reservoir must be mixing (and not, e.g., simply ergodic), in
order that no discrete eigenvalue different from $0$ exists.
Otherwise, the point spectrum (except $0$) is out of control and
again the emergence of  secular terms is inevitable. Moreover, the
mixing property of the reservoir is crucial even for a
\textit{factorized} initial state like (\ref{eqn:rhoprod}). Indeed,
careful scrutiny of the proof shown in this section shows that the
divergence of $\mathcal{K}_{mn}^{(\lambda)}(\tau)$ has nothing to do
with the initial correlation. In this sense, although our original
motivation was the study of the factorization Ansatz, the results
have more general validity. The reason why this was unnoticed so far
is the following. In the standard textbook derivations of the master
equation, a factorized initial state is assumed and the projection
(\ref{eqn:defproj}) is defined in terms of the same reference state
$\Omega_\mathrm{B}$ that is contained in the factorized initial
state (\ref{eqn:rhoprod}): in practically all examples, this is
taken to be the canonical state at temperature $T$, which is clearly
mixing, and all the criteria listed in Sec.\ \ref{sec:theoremmain}
are satisfied. However, recently, more complicated physical
situations are attracting increasing interest, such as those related
to NESS shown  in Fig.\ \ref{fig:ness}(b). These cases call for a
characterization of the state of the reservoir, that is
mixing and satisfies the hypotheses in Sec.\ \ref{sec:theoremmain} required to prove the theorem \cite{ref:Ruelle}.

\section{Factorization at All Times}
\label{sec:Qpart}
Time $t=0$ has no particular status: one can prove the same master
equation (\ref{eqn:diffeq}) with a factorized initial state for any
``initial'' time $t_0$. This means that the limiting dynamics is
such that the density matrix remains \textit{factorized at all
times}: the reservoir state does not evolve, while  the system state
follows the master equation (\ref{eqn:diffeq}). In order to show
this, one must prove the validity of Eq.\ (\ref{eqn:Theorem2}) in
van Hove's limit.
Then, clearly, no spurious term will develop in the master equation
and no correlations can appear.

By integrating (\ref{eqn:A3A}) and inserting it into
(\ref{eqn:A3B}), we get the following integral equation for
$\mathcal{Q}\rho(t)$,
\begin{eqnarray}
\mathcal{Q}\rho(t)
=\e^{\mathcal{L}_0't}\mathcal{Q}\rho_0
&+&\lambda\rint_0^t\d
t'\,\e^{\mathcal{L}_0'(t-t')}
\mathcal{Q}\mathcal{L}_\mathrm{SB}\e^{\mathcal{L}_\mathrm{S}t'}
\mathcal{P}\rho_0\nonumber\\
&+&\lambda^2\rint_0^t\d t'\rint_0^{t'}\d
t''\,\e^{\mathcal{L}_0'(t-t')}
\mathcal{Q}\mathcal{L}_\mathrm{SB}\e^{\mathcal{L}_\mathrm{S}(t'-t'')}
\mathcal{P}\mathcal{L}_\mathrm{SB}\mathcal{Q}\rho(t''),
\end{eqnarray}
which is rearranged, by interchanging the integrations in the last
term, to yield in the scaled time $\tau=\lambda^2t$
\begin{multline}
\mathcal{Q}\rho(\tau/\lambda^2)
=\mathcal{Q}\e^{\mathcal{L}_0'\tau/\lambda^2}\Biggl(
\mathcal{Q}\rho_0
-\lambda\sum_m\mathcal{R}_m^{(\lambda)}(-\tau)
\mathcal{L}_\mathrm{SB}\tilde{Q}_m\mathcal{P}\rho_0\\
{}-\sum_m\rint_0^\tau
\d\tau'\,\e^{-\mathcal{L}_0'\tau'/\lambda^2}
\mathcal{R}_m^{(\lambda)}(\tau'-\tau)
\mathcal{L}_\mathrm{SB}\tilde{Q}_m\mathcal{P}\mathcal{L}_\mathrm{SB}
\mathcal{Q}\rho(\tau'/\lambda^2)
\Biggr),
\label{eqn:Qpartscaled}
\end{multline}
where $\mathcal{R}_m^{(\lambda)}(\tau)$ is defined in
(\ref{eqn:KEY}). Under the proper choice of the projection
$\mathcal{P}$, the kernel $\mathcal{R}_m^{(\lambda)}(\tau)$ is
bounded for any $\tau$ and $\lambda$ (even for $\lambda\to0$), as
already discussed in the previous section and in Appendix
\ref{app:Rlambdatau}, and it is possible to show that (see
Appendix \ref{app:Rlambdatau})
\begin{equation}
\lim_{\lambda\to0}\mathcal{Q}\e^{\mathcal{L}_0'\tau/\lambda^2}
=0 .
\label{eqn:PropQecay}
\end{equation}
Then, the integrand in the last term in (\ref{eqn:Qpartscaled}) is
bounded, all the terms in the parentheses are finite, and the
prefactor in (\ref{eqn:Qpartscaled}) vanishes according to
(\ref{eqn:PropQecay}), which proves (\ref{eqn:Theorem2}), the second
part of the theorem in Sec.\ \ref{sec:theoremmain}.

It is worth noting that the interaction between system S and
reservoir B is not essential to the factorization; the free evolution
eliminates the correlation, and the reservoir relaxes into the
mixing state $\Omega_\mathrm{B}$. Indeed, for any state $\rho_0$ of
the total system of the type (\ref{eqn:CondInitialState}), we have
\begin{equation}
\e^{\mathcal{L}_0t}\rho_0
\tolimit{t\to\infty}{\longrightarrow}
\e^{\mathcal{L}_\mathrm{S}t}\Pi_0\rho_0
=\e^{\mathcal{L}_\mathrm{S}t}
\tr_\mathrm{B}\{\rho_0\}\otimes\Omega_\mathrm{B},
\label{eqn:FreeFactorization}
\end{equation}
where the contribution of the absolutely continuous spectrum decays
out due to Riemann--Lebesgue's lemma. [See
Eqs.\ (\ref{eq:specrestxt}) and (\ref{eqn:BathRelaxation}).] In the
rescaled time $\tau$, the factorization is very rapid, and the total
system looks factorized at any moment (if the observables that one
can measure on the reservoir are local enough: a concrete example
will be discussed in the following Article II \cite{ref:ArticleII}). Summarizing,
reservoir B relaxes into the mixing state $\Omega_\mathrm{B}$
through its own free evolution, yielding the factorization of the
state of the total system, while system S dissipates through the
interaction: a remarkable and consistent global view.

It is also interesting to compare the present result with
Bogoliubov's view on the classical gas dynamics
\cite{ref:KuboTextbook,Bogoliubov,Ford-Uhlenbeck}. According to this
view, ``molecular chaos'' erases a large amount of initial
information and the system reaches the so-called kinetic stage,
where the one-body distribution function governs the evolution of
the whole system and obeys the Boltzmann equation, irrespectively of
the initial conditions. In the present case, the initial loss of
system-reservoir correlations corresponds to the information loss
due to molecular chaos, and the stage described by the master
equation in van Hove's limit corresponds to the kinetic stage.

\section{Concluding Remarks}
\label{sec:Summary}
We analyzed the assumption of factorization of the initial state in
the dynamics of a quantum mechanical system in interaction with a
reservoir. In van Hove's limit, the dynamics can be consistently
described in terms of a master equation, but a correct application
of Nakajima--Zwanzig's projection method requires that the reference
state of the reservoir, in terms of which the projection operator is
defined, be endowed with the mixing property. If the reservoir
dynamics is not mixing, the evolution develops secular terms.
In the above discussion, one implicitly assumes that the van Hove limit (\ref{eqn:densdef}) exists and is the solution of Eqs.\ (\ref{eqn:inteq})--(\ref{eqn:diffeq}).

In the standard situation, when a small system is coupled to a
reservoir at a given temperature, the choice of the reference state
is obvious and is simply the canonical state (that is also mixing).
In more articulated situations, such as those of NESS, the choice of
the reference state requires care and need to be characterized: our
analysis shows that a correct reference state must in general be
mixing.

This Ansatz yields the standard procedure and the usual master
equation. As a byproduct, one observes that the mixing property is
crucial even when the initial state is assumed to be factorized,
otherwise the presence of a secular terms is inevitable. In this
respect our analysis, that was originally motivated by the study of
the assumption of initial state factorization, has more general
validity.

We shall see in the following Article II \cite{ref:ArticleII}, by close scrutiny of some
explicit examples, that Markovianity becomes a valid approximation
for timescales that depend both on the form factors of the
interaction and on the spatial extension of the local
observables that can be measured on the reservoir. This will
corroborate and complement the general findings discussed in this
article. Other issues, such as the spin-offs on the complete
positivity of the dynamics
\cite{ref:Pechukas,ref:LindbladLendi,ref:Royer,ref:Buzek,ref:Kraus,ref:Benatti,ref:Hanggi2004,ref:SudarshanPRA2004},
the mathematical conditions at the origin of van Hove's limit \cite{ref:Antoniou}, 
as well as the consequences of higher order corrections
\cite{ref:Agarwal} and their interplay with initial correlations
will be investigated in a future paper.

\ack We thank D.A.\ Lidar and A.\ Shabani for discussions. This work
is partly supported by the bilateral Italian--Japanese Projects
II04C1AF4E on ``Quantum Information, Computation and Communication''
of the Italian Ministry of Instruction, University and Research, and
15C1 on ``Quantum Information and Computation'' of the Italian
Ministry for Foreign Affairs, by the European Community through the Integrated Project EuroSQIP, by the Grant for The 21st Century COE
Program ``Holistic Research and Education Center for Physics of
Self-Organization Systems'' at Waseda University, the Grant-in-Aid
for the COE Research ``Establishment of Molecular Nano-Engineering
by Utilizing Nanostructure Arrays and Its Development into
Micro-Systems'' at Waseda University (No.\ 13CE2003), and the
Grants-in-Aid for Scientific Research on Priority Areas ``Control of
Molecules in Intense Laser Fields'' (No.\ 14077219), ``Dynamics of
Strings and Fields'' (No.\ 13135221), and for Young Scientists (B) (No.\ 18740250) from the Ministry of Education,
Culture, Sports, Science and Technology, Japan, and by Grants-in-Aid
for Scientific Research (C) (Nos.\ 14540280, 17540365, and 18540292) from
the Japan Society for the Promotion of Science.

\appendix
\section{Mixing Property}\label{app:mixing}
The free Liouvillian of an \textit{infinite} reservoir
$\mathcal{L}_\mathrm{B}$ has a point spectrum
\cite{ref:JaksicPillet,BR,ref:Haag}. The clue for our problem is
to handle it properly, by making use of the right projection
$\mathcal{P}$. The mixing property of the reservoir plays an
important role in this context. Let us hence briefly recall these
notions.

\subsection{Mixing Property and Spectrum of the Liouvillian}
\label{subsec:mixing}
The state $\Omega_\mathrm{B}$ is said to be \textit{mixing} with
respect to the reservoir dynamics $\e^{\mathcal{L}_\mathrm{B}t}$,
if the correlation function of any two bounded operators of the
reservoir, $X$ and $Y$, behaves as (\ref{eqn:MixingB})
\cite{ref:JaksicPillet,ref:Frohlich,BR,ref:Haag,ref:ReedSimon}. The ordinary
canonical equilibrium state of free bosons at a finite temperature
is a typical mixing state. Other interesting examples are
nonequilibrium steady states (NESS).

It is important to observe that the mixing property is strongly
related to the spectral properties of the  Liouvillian
$\mathcal{L}_\mathrm{B}$ \cite{ref:ReedSimon}. Let us consider a
reservoir state
$\rho_\mathrm{B}=\Lambda_\mathrm{B}\Omega_\mathrm{B}$
($\tr_\mathrm{B}\rho_\mathrm{B}=1$) related to a mixing state
$\Omega_\mathrm{B}$ by a bounded superoperator $\Lambda_\mathrm{B}$,
in the sense of (\ref{eqn:CondInitialState}). Then, by setting
$Y\Omega_\mathrm{B}=\Lambda_\mathrm{B}\Omega_\mathrm{B}
=\rho_\mathrm{B}$ in (\ref{eqn:MixingB}), the mixing property of
$\Omega_\mathrm{B}$ translates into
\begin{equation}
\langle X(t)\rangle_{\rho_\mathrm{B}} \to\langle
X\rangle_{\Omega_\mathrm{B}} \quad\mathrm{as}\quad t\to\infty,
\label{eqn:Relaxation}
\end{equation}
and in this sense,
\begin{equation}
\e^{\mathcal{L}_\mathrm{B}t}\rho_\mathrm{B}
\to\Omega_\mathrm{B}\quad\mathrm{as}\quad t\to\infty,
\label{eqn:RelaxationB}
\end{equation}
i.e., the state $\rho_\mathrm{B}$ relaxes towards the mixing state
$\Omega_\mathrm{B}$.

Let us consider the spectral resolution of
$\e^{\mathcal{L}_\mathrm{B}t}\rho_\mathrm{B}$,
\begin{equation}
\e^{\mathcal{L}_\mathrm{B}t}\rho_\mathrm{B}
=\sum_{\nu_j}\e^{-i\nu_jt}\Pi_j\rho_\mathrm{B}
+\rint \e^{-i\nu t}\,\d\Pi(\nu)\rho_\mathrm{B},
\label{eq:specres}
\end{equation}
where $\Pi_j$ and $\Pi(\nu)$ are the spectral projections of
$\mathcal{L}_\mathrm{B}$ belonging to its possible discrete
eigenvalues $\{\nu_j\}$ and to its absolutely continuous spectrum
$\{\nu\}$, respectively. The second term, representing the
contribution of the absolutely continuous spectrum, decays out as
$t\to\infty$ due to Riemann--Lebesgue's lemma, but the first term,
the contribution of the point spectrum, survives. Property
(\ref{eqn:RelaxationB}) means that (within the class of states of
the form $\rho_\mathrm{B}=\Lambda_\mathrm{B}\Omega_\mathrm{B}$)
there exists the only simple eigenvalue $0$ of the Liouvillian
$\mathcal{L}_\mathrm{B}$ with the eigenprojection $\Pi_0$
satisfying $\Pi_0\rho_\mathrm{B}=\Omega_\mathrm{B}$: mixing
forbids the existence of other eigenvalues than $0$, reducing
(\ref{eq:specres}) to (\ref{eq:specrestxt}), and the eigenvalue
$0$ is not degenerated within the single sector specified by
$\Omega_\mathrm{B}$. Furthermore, since
$\Pi(\nu)\Omega_\mathrm{B}=0$ [remember the orthogonality
$\Pi_0\Pi(\nu)=\Pi(\nu)\Pi_0=0$], the mixing state
$\Omega_\mathrm{B}$ is a stationary state with respect to the
reservoir dynamics $\e^{\mathcal{L}_\mathrm{B}t}$, i.e.\ Eq.\
(\ref{BathEq}).

If we only require that $0$ be a simple eigenvalue of
$\mathcal{L}_\mathrm{B}$, but we do not care about the rest of the
spectrum, then property (\ref{eqn:MixingB}) only holds in a Cesaro
sense, namely
\begin{equation}
\lim_{T\to\infty}\frac{1}{T}\rint_0^T\d t\,\langle
X(t)Y\rangle_{\Omega_\mathrm{B}} =\langle
X\rangle_{\Omega_\mathrm{B}} \langle Y\rangle_{\Omega_\mathrm{B}},
\label{eqn:ergodicityA}
\end{equation}
and $\Omega_\mathrm{B}$ is called an \textit{ergodic} state.
Ergodicity suffices to show the stationarity (\ref{BathEq}) of
$\Omega_\mathrm{B}$; the mixing property is not necessary
\cite{ref:JaksicPillet,ref:Frohlich,BR,ref:Haag,ref:ReedSimon}. The uniqueness
of the point spectrum $0$, however, is equivalent to weak mixing
\cite{ref:JaksicPillet,BR,ref:Haag,ref:ReedSimon} $\int_0^T\d t\,|
\langle X(t)Y\rangle_{\Omega_\mathrm{B}} -\langle
X\rangle_{\Omega_\mathrm{B}} \langle
Y\rangle_{\Omega_\mathrm{B}}|/T \to0$ and this implies that the
remaining part of the spectrum can also contain singular
continuous components. In such a case, one might conjecture that
the theorem of Sec.\ \ref{sec:theoremmain} is still valid, but in
a weaker sense.

\subsection{A Solvable Example}
\label{sec:solvable mixing}
Let us demonstrate the mixing property (\ref{eqn:MixingB}) in a solvable
example. Let us consider the reservoir Hamiltonian
\begin{equation}
H_\mathrm{B}=\rint\d\omega\,\omega b_\omega^\dag b_\omega
\label{eqn:ReservoirHamiltonian}
\end{equation}
and the reservoir dynamics starting from the initial state
\begin{equation}
\rho_\mathrm{B} =\tilde{\Lambda}_\mathrm{B}\rho_\mathcal{W}
=\rint\d\omega\rint\d\omega'\,
w_{\omega\omega'}b_\omega^\dag\rho_\mathcal{W}b_{\omega'}
\label{eqn:PerturbedGaussianReservoirState}
\end{equation}
with
\begin{equation}
\rho_\mathcal{W} =\frac{1}{Z_\mathcal{W}}\exp\!\left( -\rint
\d\omega\rint \d\omega'\,
b_\omega^\dag\mathcal{W}_{\omega\omega'}b_{\omega'} \right),
\label{eqn:GaussianReservoirState}
\end{equation}
where $b_\omega$ ($b_\omega^\dag$) is the annihilation (creation)
operator of the reservoir, satisfying the canonical commutation
relation $[b_\omega,b_{\omega'}^\dag]=\delta(\omega-\omega')$. The
states $\rho_\mathrm{B}$ and $\rho_\mathcal{W}$ are normalized,
$\tr_\mathrm{B}\rho_\mathrm{B}=1$ and
$\tr_\mathrm{B}\rho_\mathcal{W}=1$, with the normalization constant
$Z_\mathcal{W}$. $w_{\omega\omega'}$ is a bounded and Hermitian
positive matrix ($w_{\omega\omega'}=w_{\omega'\omega}^*$), and the
Gaussian state $\rho_\mathcal{W}$ is perturbed by the bounded
superoperator $\tilde{\Lambda}_\mathrm{B}$. Furthermore,
$\mathcal{W}_{\omega\omega'}$ is Hermitian
($\mathcal{W}_{\omega\omega'}=\mathcal{W}_{\omega'\omega}^*$) and
consists of $\mathcal{W}_{\omega\omega'}^{(0)}$, that is
proportional to $\delta(\omega-\omega')$, and the remaining part
$\tilde{\mathcal{W}}_{\omega\omega'}$,
\begin{equation}
\mathcal{W}_{\omega\omega'} =\mathcal{W}_{\omega\omega'}^{(0)}
+\tilde{\mathcal{W}}_{\omega\omega'},\qquad
\mathcal{W}_{\omega\omega'}^{(0)}=W(\omega)\delta(\omega-\omega').
\label{eqn:Wdecomp}
\end{equation}

The Gaussian state $\rho_\mathcal{W}$ is fully characterized by the
two-point function
\begin{equation}
\mathcal{N}_{\omega\omega'} =\langle b_{\omega'}^\dag
b_\omega\rangle_{\rho_\mathcal{W}}
=N(\omega)\delta(\omega-\omega')
+\tilde{\mathcal{N}}_{\omega\omega'},
\label{eqn:ExtBoseFunc}
\end{equation}
where the first term, proportional to $\delta(\omega-\omega')$, is
the expectation value of the number operator in the state
$\rho_{\mathcal{W}_0}$,
\begin{equation}
\mathcal{N}_{\omega\omega'}^{(0)} =\langle b_{\omega'}^\dag
b_\omega\rangle_{\rho_{\mathcal{W}_0}}
=N(\omega)\delta(\omega-\omega'),\qquad
N(\omega)=\frac{1}{\e^{W(\omega)}-1}
\label{eqn:BoseFunc}
\end{equation}
with
\begin{equation}
\rho_{\mathcal{W}_0} =\frac{1}{Z_{\mathcal{W}_0}}\exp\!\left(
-\rint\d\omega\,b_\omega^\dag W(\omega)b_\omega \right),\qquad
\tr_\mathrm{B}\rho_{\mathcal{W}_0}=1.
\label{eqn:NonlocalGaussianState}
\end{equation}
In fact, $\rho_\mathcal{W}$ is different from $\rho_{\mathcal{W}_0}$
by a bounded operator $L_\mathrm{B}$, as
\begin{equation}
\rho_\mathcal{W}
=L_\mathrm{B}\rho_{\mathcal{W}_0}L_\mathrm{B}^\dag,
\end{equation}
where
\begin{equation}
L_\mathrm{B} =\rho_\mathcal{W}^{1/2}\rho_{\mathcal{W}_0}^{-1/2}
=\sqrt{\frac{Z_{\mathcal{W}_0}}{Z_\mathcal{W}}}
\mathop{\bar{\mathrm{T}}}\exp\Biggl( -\rint_0^{1/2}\d\beta\,
b^\dag \e^{-\beta\mathcal{W}^{(0)}}\tilde{\mathcal{W}}
\e^{\beta\mathcal{W}^{(0)}}b \Biggr) ,
\label{eqn:LB}
\end{equation}
with $\mathop{\bar{\mathrm{T}}}$ denoting the anti-chronologically
ordered product and
\begin{equation}
b^\dag\e^{-\beta\mathcal{W}^{(0)}}\tilde{\mathcal{W}}
\e^{\beta\mathcal{W}^{(0)}}b =\rint\d\omega\rint\d\omega'\,
b_\omega^\dag\e^{-\beta
W(\omega)}\tilde{\mathcal{W}}_{\omega\omega'} \e^{\beta
W(\omega')}b_{\omega'}.
\end{equation}
Hence, by applying Wick's theorem, the two-point function
(\ref{eqn:ExtBoseFunc}) reads
\begin{eqnarray}
\mathcal{N}_{\omega\omega'}
&=&\langle L_\mathrm{B}^\dag b_{\omega'}^\dag b_\omega
L_\mathrm{B}\rangle_{\rho_{\mathcal{W}_0}} = \langle
b_{\omega'}^\dag b_\omega\rangle_{\rho_{\mathcal{W}_0}}
\langle L_\mathrm{B}^\dag L_\mathrm{B}\rangle_{\rho_{\mathcal{W}_0}}
+\cdots.
\end{eqnarray}
The first term is $\mathcal{N}_{\omega\omega'}^{(0)}$ in
(\ref{eqn:BoseFunc}), since $\langle L_\mathrm{B}^\dag
L_\mathrm{B}\rangle_{\rho_{\mathcal{W}_0}}=\tr_\mathrm{B}\rho_\mathcal{W}=1$,
and the other terms, defining
$\tilde{\mathcal{N}}_{\omega\omega'}$, are bounded functions, not
proportional to $\delta(\omega-\omega')$.

Let us now take any two bounded operators of the reservoir, of the
form
\begin{equation}
X=\rint\d\omega\rint\d\omega'\,
b_\omega^\dag\mathcal{X}_{\omega\omega'}b_{\omega'},\qquad
Y=\rint\d\omega\rint\d\omega'\,
b_\omega^\dag\mathcal{Y}_{\omega\omega'}b_{\omega'},
\label{eqn:LocalOps}
\end{equation}
and observe how the mixing property (\ref{eqn:MixingB}) emerges. In
this case, the correlation function reads
\begin{eqnarray}
\langle X(t)Y\rangle_{\rho_\mathrm{B}}
&=&\langle \e^{i
H_\mathrm{B}t}X\e^{-i H_\mathrm{B}t}Y
\rangle_{\rho_\mathrm{B}}\nonumber\\
&=&\rint\d\omega_1\rint\d\omega_2\rint\d\omega_3\rint\d\omega_4
\rint\d\omega\rint\d\omega'\,
\mathcal{X}_{\omega_1\omega_2}\mathcal{Y}_{\omega_3\omega_4}
w_{\omega\omega'}\nonumber\\
&&\times\langle b_{\omega'}b_{\omega_1}^\dag
b_{\omega_2}b_{\omega_3}^\dag b_{\omega_4}b_\omega^\dag
\rangle_{\rho_\mathcal{W}}\e^{i(\omega_1-\omega_2)t}.
\label{eqn:Correlation}
\end{eqnarray}
By applying Wick's theorem,
$\langle b_{\omega'}b_{\omega_1}^\dag b_{\omega_2}b_{\omega_3}^\dag
b_{\omega_4}b_\omega^\dag\rangle_{\rho_\mathcal{W}}$
can be expressed in terms of two-point functions
(\ref{eqn:ExtBoseFunc}), and by applying Riemann--Lebesgue's lemma,
only the contribution of $\delta(\omega_1-\omega_2)$ in $\langle
b_{\omega_1}^\dag b_{\omega_2}\rangle_{\rho_\mathcal{W}}$ survives
in the long-time limit, to yield
\begin{eqnarray}
\langle X(t)Y\rangle_{\rho_\mathrm{B}}
&\tolimit{t\to\infty}{\longrightarrow}&\rint\d\omega_1\rint\d\omega_3\rint\d\omega_4
\rint\d\omega\rint\d\omega'\,
\mathcal{X}_{\omega_1\omega_1}\mathcal{Y}_{\omega_3\omega_4}
w_{\omega\omega'}\nonumber\\
&&{}\times N(\omega_1)\langle b_{\omega'}b_{\omega_3}^\dag
b_{\omega_4} b_\omega^\dag
\rangle_{\rho_\mathcal{W}}
\nonumber\\
&=&\langle X\rangle_{\rho_{\mathcal{W}_0}}
\langle Y\rangle_{\rho_\mathrm{B}}.
\label{eqn:MixingGeneral}
\end{eqnarray}
Therefore, if $\rho_\mathrm{B}=\rho_{\mathcal{W}_0}$ (i.e.,
$\tilde{\mathcal{W}}_{\omega\omega'}=0$ without the perturbation
$\tilde{\Lambda}_\mathrm{B}$), Eq.\ (\ref{eqn:MixingGeneral}) is
nothing but the definition of mixing in (\ref{eqn:MixingB}), and
$\rho_{\mathcal{W}_0}$ given in (\ref{eqn:NonlocalGaussianState}) is
an example of mixing state. The canonical state
$\rho_{\mathcal{W}_0}$ with $W(\omega)=\beta\omega$, is thus a
typical mixing state. We have demonstrated (\ref{eqn:MixingGeneral})
with the specific observables $X$ and $Y$ in (\ref{eqn:LocalOps}),
but this example helps us understand how mixing emerges for general
observables.

It is important to note that, in this appendix, we have considered
only the reservoir dynamics generated by the reservoir Hamiltonian
$H_\mathrm{B}$, without any interaction. The free evolution is
responsible for the mixing.

The mixing property (\ref{eqn:MixingB}) is demonstrated here with
the thermal equilibrium state (\ref{eqn:NonlocalGaussianState}).
It is also possible to prove it for the NESS depicted in Fig.\
\ref{fig:ness}(b). See Ref.\ \cite{ref:Ruelle} for details.

\section{Inequivalent Sectors}\label{app:InequivSec}
The main purpose of this tutorial appendix is to clarify that
different mixing states belong to different sectors which are
inequivalent to each other, and any bounded perturbation on a mixing
state does not bring it to a different sector. Let us demonstrate
the inequivalence of the sectors with an explicit example that
captures the essence of the inequivalent representation.

In order to analyze an infinitely extended system, let us begin with
a free bosonic gas in a 1D box whose size is specified by a
parameter $\ell$, and then take the continuum limit $\ell\to\infty$.
We consider two canonical states $\Omega_\beta$ and
$\Omega_{\beta'}$ with different temperatures and compute the
overlap between them through the quantity
$\tr_\mathrm{B}\{\Omega_\beta^{1/2}\Omega_{\beta'}^{1/2}\}$. Note
that the canonical state is a mixing state as shown in
Appendix \ref{app:mixing}\@.

In the finite box, momentum $k$ is discrete, and the Hamiltonian of
the bosonic gas is given by
\begin{equation}
H_\mathrm{B}=\sum_k\omega_k{b_k^{(\ell)}}^\dag b_k^{(\ell)},
\end{equation}
where $b_k^{(\ell)}$ and ${b_k^{(\ell)}}^\dag$ satisfy the
canonical commutation relation
$[b_k^{(\ell)},{b_{k'}^{(\ell)}}^\dag]=\delta_{kk'}$. The
canonical state at the inverse temperature $\beta$ is given by
\begin{equation}
\Omega_\beta=\frac{1}{Z_\beta}\e^{-\beta H_\mathrm{B}},\qquad
Z_\beta^{-1}=\prod_k(1-\e^{-\beta\omega_k}),
\label{eqn:DiscreteCanonicalState}
\end{equation}
and the overlap between $\Omega_\beta$ and $\Omega_{\beta'}$ reads
\begin{eqnarray}
\tr_\mathrm{B}\{\Omega_\beta^{1/2}\Omega_{\beta'}^{1/2}\}
&=&\prod_k\frac{\sqrt{(1-\e^{-\beta\omega_k})
(1-\e^{-\beta'\omega_k})}}{1-\e^{-\bar{\beta}\omega_k}}\nonumber\\
&=&\exp\!\left[
-\frac{1}{2}\sum_k\ln\frac{(1-\e^{-\bar{\beta}\omega_k})^2}%
{(1-\e^{-\beta\omega_k})(1-\e^{-\beta'\omega_k})} \right],
\label{eqn:OverlapFinite}
\end{eqnarray}
where $\bar{\beta}=(\beta+\beta')/2$. The exponent in
(\ref{eqn:OverlapFinite}) is easily shown to be less than zero for
$\beta\neq\beta'$ and equal to zero for $\beta=\beta'$. In the
continuum limit $\ell\to\infty$, the summation in the exponent is
replaced with an integral as $\sum_k\to(\ell/2\pi)\rint \d k$, and
the overlap is reduced to
\begin{eqnarray}
\tr_\mathrm{B}\{\Omega_\beta^{1/2}\Omega_{\beta'}^{1/2}\}
&=&\exp\!\left[ -\frac{\ell}{4\pi}\rint\d k
\ln\frac{(1-\e^{-\bar{\beta}\omega_k})^2}%
{(1-\e^{-\beta\omega_k})(1-\e^{-\beta'\omega_k})}
\right]\nonumber\\
&\to&
\begin{cases}
1&(\beta=\beta')\\
0&(\beta\neq\beta')
\end{cases}
\quad\mathrm{as}\quad\ell\to\infty,
\label{eqn:Inequivalence}
\end{eqnarray}
which means that canonical states with different temperatures do not
overlap and belong to inequivalent sectors.

Any bounded perturbation does not change the situation: a state
$\tilde{\Omega}_\beta$, which is different from a canonical state
$\Omega_\beta$ only by a bounded superoperator, belongs to the same
sector as that of the canonical state $\Omega_\beta$ and does not
overlap with a canonical state $\Omega_{\beta'}$ with different
temperature. Consider, for example, a state $\tilde{\Omega}_\beta$,
whose square root $\tilde{\Omega}_\beta^{1/2}$ is different from
$\Omega_\beta^{1/2}$ by a bounded operator $K_\mathrm{B}$ (or
$L_\mathrm{B}$) as
\begin{equation}
\tilde{\Omega}_\beta^{1/2}
=K_\mathrm{B}\Omega_\beta^{1/2}K_\mathrm{B}^\dag
=L_\mathrm{B}\Omega_\beta^{1/2}
=\Omega_\beta^{1/2}L_\mathrm{B}^\dag,\quad
L_\mathrm{B}=K_\mathrm{B}\Omega_\beta^{1/2}K_\mathrm{B}^\dag\Omega_\beta^{-1/2}
\end{equation}
with the normalization conditions
$\tr_\mathrm{B}\tilde{\Omega}_\beta=1$ and
$\tr_\mathrm{B}\Omega_\beta=1$. We again begin with a finite $\ell$,
so that the overlap between $\tilde{\Omega}_\beta$ and
$\Omega_{\beta'}$ now reads
\begin{equation}
\tr_\mathrm{B}\{\tilde{\Omega}_\beta^{1/2}\Omega_{\beta'}^{1/2}\}
=\tr_\mathrm{B}\{
L_\mathrm{B}\Omega_\beta^{1/2}\Omega_{\beta'}^{1/2} \}
=\prod_k\frac{\sqrt{(1-\e^{-\beta\omega_k})
(1-\e^{-\beta'\omega_k})}}{1-\e^{-\bar{\beta}\omega_k}} \langle
L_\mathrm{B}\rangle_{\bar{\beta}},
\label{eqn:OverlapFiniteLocal}
\end{equation}
where $\langle L_\mathrm{B}\rangle_{\bar{\beta}}$ is the expectation
value of $L_\mathrm{B}$ in the canonical state at temperature
$\bar{\beta}$. Note that $\langle L_\mathrm{B}\rangle_{\bar{\beta}}$
is finite even in the continuum limit $\ell\to\infty$, since
$L_\mathrm{B}$ is a bounded operator. Therefore, exactly the same
argument as (\ref{eqn:Inequivalence}) applies to this case and leads
to the conclusion
\begin{equation}
\tr_\mathrm{B}\{\tilde{\Omega}_\beta^{1/2}\Omega_{\beta'}^{1/2}\}
\to
\begin{cases}
\langle L_\mathrm{B}\rangle_\beta&(\beta=\beta')\\
0&(\beta\neq\beta')
\end{cases}
\quad\mathrm{as}\quad\ell\to\infty,
\end{equation}
i.e., $\tilde{\Omega}_\beta$ does not overlap with $\Omega_{\beta'}$
and belongs to the sector equivalent to $\Omega_\beta$.

\section{Diagonal Projection}\label{app:DiagonalProjection}
Let us confirm the property of the diagonal projection in
(\ref{eqn:DiagonalProjectionLocal}) with an explicit example. We
consider the same model as in Appendix \ref{app:InequivSec} and
observe how the diagonal projection acts on a reservoir state
$\tilde{\Omega}_\beta=\Lambda_\mathrm{B}\Omega_\beta$, which is
different from the canonical state $\Omega_\beta$ at the inverse
temperature $\beta$ only by a bounded superoperator
$\Lambda_\mathrm{B}$, and therefore belongs to the sector equivalent
to $\Omega_\beta$.

We begin with a 1D bosonic gas in a finite box, and then take the
continuum limit $\ell\to\infty$. For a finite $\ell$, the diagonal
projection $\mathcal{P}_\mathrm{D}$ is defined as
(\ref{eqn:DiagProjDiscrete}) and is given in this case by
\begin{equation}
\mathcal{P}_\mathrm{D}\tilde{\Omega}_\beta
=\sum_{\{n_k\}}\ket{\{n_k\}}\bra{\{n_k\}}
\tilde{\Omega}_\beta\ket{\{n_k\}}\bra{\{n_k\}},
\end{equation}
where $\ket{\{n_k\}}=\ket{n_{k_1}n_{k_2}\dots}$, $n_k$ being the
occupation number in mode $\omega_k$. To be explicit, let us take a
reservoir state
\begin{equation}
\tilde{\Omega}_\beta =\Lambda_\mathrm{B}\Omega_\beta
=\sum_{k,k'}w_{kk'}^{(\ell)}{b_k^{(\ell)}}^\dag \Omega_\beta
b_{k'}^{(\ell)} =\sum_{k,k'}w_{kk'}^{(\ell)}\e^{\beta\omega_{k'}}
{b_k^{(\ell)}}^\dag b_{k'}^{(\ell)}\Omega_\beta
\label{eqn:PerturbedCanonicalState}
\end{equation}
with a Hermitian ($w_{kk'}^{(\ell)}={w_{k'k}^{(\ell)}}^*$) positive
matrix, and consider the expectation value of a bounded operator of
the reservoir
\begin{equation}
Y=\sum_{k,k'}\mathcal{Y}_{kk'}^{(\ell)}{b_k^{(\ell)}}^\dag
b_{k'}^{(\ell)}
\label{eqn:ReservoirVarableFinite}
\end{equation}
in the projected state $\mathcal{P}_\mathrm{D}\tilde{\Omega}_\beta$:
\begin{eqnarray}
\tr_\mathrm{B}\{Y\mathcal{P}_\mathrm{D}\tilde{\Omega}_\beta\}
&=&\sum_{\{n_k\}}\bra{\{n_k\}}Y\ket{\{n_k\}}\bra{\{n_k\}}
\tilde{\Omega}_\beta\ket{\{n_k\}}\nonumber\\
&=&\sum_{\{n_k\}}\bra{\{n_k\}}Y\ket{\{n_k\}}
\bra{\{n_k\}}L_\mathrm{B}\ket{\{n_k\}}
\bra{\{n_k\}}\Omega_\beta\ket{\{n_k\}},
\label{eqn:ExpectValueProjected}
\end{eqnarray}
where $L_\mathrm{B}=\sum_{k,k'}w_{kk'}^{(\ell)}\e^{\beta\omega_{k'}}
{b_k^{(\ell)}}^\dag b_{k'}^{(\ell)}$ is the bounded operator acting
on the left side of $\Omega_\beta$ in
(\ref{eqn:PerturbedCanonicalState}). Note that $\Omega_\beta$ is
diagonal with respect to the basis $\ket{\{n_k\}}$ and the diagonal
elements are given by
\begin{equation}
\bra{\{n_k\}}\Omega_\beta\ket{\{n_k\}}
=\frac{1}{Z_\beta}\e^{-\beta\sum_kn_k\omega_k}
\end{equation}
with $Z_\beta$ given in (\ref{eqn:DiscreteCanonicalState}). The
diagonal elements of $Y$ and $L_\mathrm{B}$ are easily evaluated to
be
\begin{equation}
\bra{\{n_k\}}Y\ket{\{n_k\}}
=\sum_k\mathcal{Y}_{kk}^{(\ell)}n_k,\quad
\bra{\{n_k\}}L_\mathrm{B}\ket{\{n_k\}}
=\sum_kw_{kk}^{(\ell)}\e^{\beta\omega_k}n_k,
\end{equation}
respectively, and Eq.\ (\ref{eqn:ExpectValueProjected}) is
\begin{equation}
\tr_\mathrm{B}\{Y\mathcal{P}_\mathrm{D}\tilde{\Omega}_\beta\}
=\frac{1}{Z_\beta}\sum_{\{n_k\}}\sum_{k,k'}
\mathcal{Y}_{kk}^{(\ell)}w_{k'k'}^{(\ell)}
\e^{\beta\omega_{k'}}n_kn_{k'}
\e^{-\beta\sum_{k''}n_{k''}\omega_{k''}}.
\label{eqn:ExpectValueProjected2}
\end{equation}
Notice here that
\begin{equation}
\langle n_kn_{k'}\rangle_\beta
=\frac{1}{Z_\beta}\sum_{\{n_k\}}n_kn_{k'}
\e^{-\beta\sum_{k''}n_{k''}\omega_{k''}} =
\begin{cases}
\langle n_k\rangle_\beta\langle n_{k'}\rangle_\beta&(k\neq k')\\
\langle n_k^2\rangle_\beta&(k=k'),
\end{cases}
\end{equation}
where
\begin{equation}
\langle n_k\rangle_\beta =\langle {b_k^{(\ell)}}^\dag
b_k^{(\ell)}\rangle_\beta =\frac{1}{\e^{\beta\omega_k}-1},\quad
\langle n_k^2\rangle_\beta =\langle({b_k^{(\ell)}}^\dag
b_k^{(\ell)})^2\rangle_\beta =2\langle n_k\rangle_\beta^2+\langle
n_k\rangle_\beta,
\end{equation}
so that Eq.\ (\ref{eqn:ExpectValueProjected2}) is decomposed into two
terms,
\begin{equation}
\tr_\mathrm{B}\{Y\mathcal{P}_\mathrm{D}\tilde{\Omega}_\beta\} =\sum_{k\neq
k'}\mathcal{Y}_{kk}^{(\ell)}w_{k'k'}^{(\ell)}
\e^{\beta\omega_{k'}}\langle n_k\rangle_\beta \langle
n_{k'}\rangle_\beta
+\sum_k\mathcal{Y}_{kk}^{(\ell)}w_{kk}^{(\ell)}
\e^{\beta\omega_{k}}\langle n_k^2\rangle_\beta.
\label{eqn:ExpectValueProjected3}
\end{equation}
We are now in a position to take the continuum limit $\ell\to\infty$
by recalling the correspondence
\begin{equation}
\sum_k\leftrightarrow\frac{\ell}{2\pi}\rint \d k,\quad
b_k^{(\ell)}\leftrightarrow\sqrt{\frac{2\pi}{\ell}}b_k,\quad
\mathcal{Y}_{kk'}^{(\ell)}\leftrightarrow\frac{2\pi}{\ell}
\mathcal{Y}_{kk'},\quad
w_{kk'}^{(\ell)}\leftrightarrow\frac{2\pi}{\ell}w_{kk'},
\end{equation}
where $b_k$ and $b_k^\dag$ satisfy the canonical commutation
relation $[b_k,b_{k'}^\dag]=\delta(k-k')$, and $\mathcal{Y}_{kk'}$
and $w_{kk'}$ are assumed to be bounded functions. The relevant
quantity now becomes
\begin{multline}
\tr_\mathrm{B}\{Y\mathcal{P}_\mathrm{D}\tilde{\Omega}_\beta\} =\rint \d
k\rint \d k'\,\mathcal{Y}_{kk} w_{k'k'}\e^{\beta\omega_{k'}}
\langle n_k\rangle_\beta\langle n_{k'}\rangle_\beta\\
+\frac{2\pi}{\ell}\rint \d k\,\mathcal{Y}_{kk}w_{kk}
\e^{\beta\omega_{k}}\langle n_k^2\rangle_\beta
\end{multline}
for large $\ell$, and the second term disappears in the continuum
limit $\ell\to\infty$. We finally obtain
\begin{equation}
\tr_\mathrm{B}\{Y\mathcal{P}_\mathrm{D}\tilde{\Omega}_\beta\} \to\langle
Y\rangle_\beta\langle L_\mathrm{B}\rangle_\beta =\langle
Y\rangle_\beta\tr_\mathrm{B}\tilde{\Omega}_\beta
\quad\mathrm{as}\quad\ell\to\infty,
\end{equation}
which yields the formula for the diagonal projection,
\begin{equation}
\mathcal{P}_\mathrm{D}\tilde{\Omega}_\beta=\Omega_\beta.
\end{equation}

The action of the diagonal projection $\mathcal{P}_\mathrm{D}$ on
the total system is now readily understood. Consider, for example, a
state of the total system
\begin{equation}
\rho =\Lambda(\openone_\mathrm{S}\otimes\Omega_\beta)
=\sum_{i,j}\rint\d k\rint \d k'\,w_{ij,kk'}S_ib_k^\dag
(\sigma_\mathrm{S}\otimes\Omega_\beta)b_{k'}S_j^\dag,
\end{equation}
where $\sigma_\mathrm{S}$ is any positive operator of system S,
$S_i$'s are system operators, and $\Lambda$ is a bounded
superoperator acting on $\openone_\mathrm{S}\otimes\Omega_\beta$.
Take an operator of the system, $A$, and an operator of the
reservoir, $Y$ given in (\ref{eqn:ReservoirVarableFinite}). Starting
with a finite $\ell$, the expectation value of the operator
$D=A\otimes Y$ in the projected state $\mathcal{P}_\mathrm{D}\rho$
is
\begin{eqnarray}
\tr\{D\mathcal{P}_\mathrm{D}\rho\}
&=&\sum_{\{n_k\}}\bra{\{n_k\}}Y\ket{\{n_k\}}\bra{\{n_k\}}
\tr_\mathrm{S}\{A\rho\}\ket{\{n_k\}}\nonumber\\
&=&\sum_{i,j}\sum_{k,k'}
\tr_\mathrm{S}\{AS_i\sigma_\mathrm{S}S_j^\dag\}
\mathcal{Y}_{kk}^{(\ell)}
w_{ij,k'k'}^{(\ell)}\e^{\beta\omega_{k'}} \langle
n_kn_{k'}\rangle_\beta,
\end{eqnarray}
which is reduced, in the continuum limit $\ell\to\infty$, to
\begin{eqnarray}
\tr\{D\mathcal{P}_\mathrm{D}\rho\} &\to&\sum_{i,j}\rint \d k\rint \d k'\,
\tr_\mathrm{S}\{AS_i\sigma_\mathrm{S}S_j^\dag\}
\mathcal{Y}_{kk}w_{ij,k'k'}\e^{\beta\omega_{k'}} \langle
n_k\rangle_\beta\langle n_{k'}\rangle_\beta
\nonumber\\
&=&\tr_\mathrm{S}[A\tr_\mathrm{B}\{\Lambda(\openone_\mathrm{S}\otimes\Omega_\beta)\}]
\langle Y\rangle_\beta\nonumber\\
&=&\tr[D(\tr_\mathrm{B}\{\rho\}\otimes\Omega_\beta)],
\end{eqnarray}
reproducing (\ref{eqn:DiagonalProjectionLocal}).

\section{Key Formulas for the Theorem}
\label{app:Rlambdatau}
Here we prove the key formulas (\ref{eqn:KeyFormula}) (with its
counterpart for $\tau<0$, which is necessary in
Sec.\ \ref{sec:Qpart}) and (\ref{eqn:PropQecay}), and see how the
proper choice of the projection $\mathcal{P}$ is crucial.

\subsection{The van Hove Limit of $\mathcal{R}_m^{(\lambda)}(\tau)$}
Let us analyze the kernel $\mathcal{R}_m^{(\lambda)}(\tau)$, defined
in (\ref{eqn:KEY}). In this appendix, the eigenprojection
$1_\mathrm{S}\otimes\Pi_0$ is written simply as $\Pi_0$.

We start by noting that in van Hove's limit, for
$n>2$,
\begin{equation}
\lambda^n\mathcal{R}_m^{(\lambda)}(\tau)
=\lambda^n\rint_0^{\tau/\lambda^2}\d t\,\mathcal{Q}
\e^{(\mathcal{L}_0'+i\omega_m)t}
\to0\quad\mathrm{as}\quad\lambda\to0\quad(n>2),
\label{eqn:Highest}
\end{equation}
\textit{irrespectively of the spectrum of $\mathcal{L}_0'$}.

Second, the following observation will be important: the convolution
\begin{equation}
\rint_0^t\d t'\,\e^{\mathcal{L}_0(t-t')}
\mathcal{Q}\mathcal{L}_\mathrm{SB}
\mathcal{Q}\e^{\mathcal{L}_0't'}
\label{eqn:Convolution}
\end{equation}
is bounded for any $t$, \textit{provided the point spectrum of
$\mathcal{L}_0$ is removed by the projection
$\mathcal{Q}=1-\mathcal{P}$ with $\mathcal{P}=\Pi_0$}. Let us look
at the Laplace transform of this convolution (for $t>0$),
\begin{equation}
\frac{1}{s-\mathcal{L}_0}\mathcal{Q}\mathcal{L}_\mathrm{SB}
\mathcal{Q}\frac{1}{s-\mathcal{L}_0'}.
\end{equation}
Neither $1/(s-\mathcal{L}_0)$ nor $1/(s-\mathcal{L}_0')$ has a
singularity on the right half plane $\Re s>0$. If $\mathcal{L}_0$
and $\mathcal{L}_0'$ have common eigenvalues (along the imaginary
axis $\Re s=0$) that are not projected out, these would give second
order poles and yield linearly diverging functions of $t$ (for large
$t$) after the inverse Laplace transform; otherwise, the convolution
decays or just oscillates. On the other hand, if the point spectrum
of $\mathcal{L}_0$ is removed by the projection $\mathcal{Q}$, such
a coincidence between the point spectra does not happen and the
convolution (\ref{eqn:Convolution}) is bounded for $t\to\infty$,
irrespectively of the point spectrum of $\mathcal{L}_0'$.

Now by using
\begin{equation}
\e^{\mathcal{L}_0't} =\e^{\mathcal{L}_0t} +\lambda\rint_0^t\d
t'\,\e^{\mathcal{L}_0(t-t')}\mathcal{Q}
\mathcal{L}_\mathrm{SB}\mathcal{Q}\e^{\mathcal{L}_0't'},
\label{eqn:Perturbation}
\end{equation}
we expand the relevant quantity (\ref{eqn:KEY}) as
\begin{equation}
\mathcal{R}_m^{(\lambda)}(\tau) =\rint_0^{\tau/\lambda^2}\d
t\,\mathcal{Q} \e^{(\mathcal{L}_0+i\omega_m)t}
+\lambda\rint_0^{\tau/\lambda^2}\d t\rint_0^t\d t'\,
\e^{(\mathcal{L}_0+i\omega_m)t}\e^{-\mathcal{L}_0t'}\mathcal{Q}
\mathcal{L}_\mathrm{SB}\mathcal{Q}\e^{\mathcal{L}_0't'}.
\label{eqn:Expansion}
\end{equation}
The first term is decomposed into two parts by the projections
$\Pi_0$ and $\Pi_\mathrm{c}=1-\Pi_0$, and the integrations are
easily carried out to give (for $\tau>0$)
\begin{eqnarray}
\rint_0^{\tau/\lambda^2}\d t\,\mathcal{Q}
\e^{(\mathcal{L}_0+i\omega_m)t} &=&\rint_0^{\tau/\lambda^2}\d
t\,\mathcal{Q}\Pi_0\e^{(\mathcal{L}_\mathrm{S}+i\omega_m)t}
+\rint_0^{\tau/\lambda^2}\d t\,\mathcal{Q}\Pi_\mathrm{c}
\e^{(\mathcal{L}_0+i\omega_m)t}\nonumber\\
&=&\sum_n\rint_0^{\tau/\lambda^2}\d t\,\mathcal{Q}\Pi_0
\tilde{Q}_n\e^{i(\omega_m-\omega_n)t} +\mathcal{Q}\Pi_\mathrm{c}
\frac{\e^{(\mathcal{L}_0+i\omega_m)\tau/\lambda^2}-1}%
{\mathcal{L}_0+i\omega_m-0^+}\nonumber\\
&=&\frac{\tau}{\lambda^2}\mathcal{Q}\Pi_0\tilde{Q}_m +\sum_{n\neq
m}\mathcal{Q}\Pi_0\tilde{Q}_n
\frac{\e^{i(\omega_m-\omega_n)\tau/\lambda^2}-1}%
{i(\omega_m-\omega_n)}\nonumber\\
&&{}+\mathcal{Q}\Pi_\mathrm{c}
\frac{\e^{(\mathcal{L}_0+i\omega_m)\tau/\lambda^2}-1}%
{\mathcal{L}_0+i\omega_m-0^+},
\label{eqn:ContPointSpec}
\end{eqnarray}
which shows that the only possible divergence of the relevant
operator in (\ref{eqn:KEY}) in van Hove's limit $\lambda\to0$
\textit{stems from the point spectrum of $\mathcal{L}_0$} (i.e.\ the
first term of the last expression). This divergence results in the
divergences of both the memory kernel
$\mathcal{K}_{mn}^{(\lambda)}(\tau)$ and the initial correlation
$\mathcal{I}^{(\lambda)}(\tau)$. \textit{However, if
$\mathcal{P}=\Pi_0$}, the divergent term disappears due to
$\mathcal{Q}\Pi_0=0$, and we have (for $\tau>0$)
\begin{equation}
\rint_0^{\tau/\lambda^2}\d t\,\mathcal{Q}
\e^{(\mathcal{L}_0+i\omega_m)t}
=\mathcal{Q}\frac{\e^{(\mathcal{L}_0+i\omega_m)\tau/\lambda^2}-1}%
{\mathcal{L}_0+i\omega_m-0^+}
\to-\frac{\mathcal{Q}}{\mathcal{L}_0+i\omega_m-0^+}
\quad\mathrm{as}\quad\lambda\to0,
\label{eqn:Zeroth}
\end{equation}
by noting the formula
\begin{equation}
\lim_{t\to\pm\infty}
\Pi_\mathrm{c}\frac{\e^{(\mathcal{L}_0+i\omega_m)t}}%
{\mathcal{L}_0+i\omega_m\mp0^+}=0 ,
\end{equation}
which is valid in the sense of distributions. The second term in
(\ref{eqn:Expansion}) can be manipulated to yield
\begin{eqnarray}
\lefteqn{\lambda\rint_0^{\tau/\lambda^2}\d t\rint_0^t\d t'\,
\e^{(\mathcal{L}_0+i\omega_m)t}\e^{-\mathcal{L}_0t'}\mathcal{Q}
\mathcal{L}_\mathrm{SB}\mathcal{Q}\e^{\mathcal{L}_0't'}
}\nonumber\\
&=&\lambda\rint_0^{\tau/\lambda^2}\d t'
\rint_{t'}^{\tau/\lambda^2}\d t\,\e^{(\mathcal{L}_0+i\omega_m)t}
\e^{-\mathcal{L}_0t'}\mathcal{Q}\mathcal{L}_\mathrm{SB}\mathcal{Q}
\e^{\mathcal{L}_0't'}\nonumber\\
&=&\lambda\frac{\mathcal{Q}}{\mathcal{L}_0+i\omega_m-0^+}
\e^{i\omega_m\tau/\lambda^2} \rint_0^{\tau/\lambda^2}\d t\,
\e^{\mathcal{L}_0(\tau/\lambda^2-t)}\mathcal{Q}
\mathcal{L}_\mathrm{SB}\mathcal{Q}\e^{\mathcal{L}_0't}
\nonumber\\
&&{}-\lambda\frac{\mathcal{Q}}{\mathcal{L}_0+i\omega_m-0^+}
\mathcal{L}_\mathrm{SB}\rint_0^{\tau/\lambda^2}\d t\,\mathcal{Q}
\e^{(\mathcal{L}_0'+i\omega_m)t}.
\label{eqn:HigherTerm}
\end{eqnarray}
The integral in the first term is the convolution
(\ref{eqn:Convolution}). Since the point spectrum of $\mathcal{L}_0$
is removed by the projection $\mathcal{Q}$, this convolution is
bounded for $\tau/\lambda^2\to\infty$.

In summary, \textit{with the choice of the projection
$\mathcal{P}=\Pi_0$}, Eq.\ (\ref{eqn:Expansion}) is arranged into the
recurrence formula
\begin{eqnarray}
\mathcal{R}_m^{(\lambda)}(\tau) &=&\mathcal{Q}
\frac{\e^{(\mathcal{L}_0+i\omega_m)\tau/\lambda^2}-1}%
{\mathcal{L}_0+i\omega_m-0^+}\nonumber\\
&&{}+\lambda\frac{\mathcal{Q}}{\mathcal{L}_0+i\omega_m-0^+}
\e^{i\omega_m\tau/\lambda^2} \rint_0^{\tau/\lambda^2}\d t\,
\e^{\mathcal{L}_0(\tau/\lambda^2-t)}\mathcal{Q}
\mathcal{L}_\mathrm{SB}\mathcal{Q}\e^{\mathcal{L}_0't}\nonumber\\
&&{}-\lambda\frac{\mathcal{Q}}{\mathcal{L}_0+i\omega_m-0^+}
\mathcal{L}_\mathrm{SB}\mathcal{R}_m^{(\lambda)}(\tau)\quad(\tau>0),
\label{eqn:Recurrence}
\end{eqnarray}
where the first term converges to (\ref{eqn:Zeroth}) and the second
term vanishes in van Hove's limit. Therefore, by iterating the above
expansion twice, we arrive at
\begin{eqnarray}
\lim_{\lambda\to0}\mathcal{R}_m^{(\lambda)}(\tau)
&=&{-\frac{\mathcal{Q}}{\mathcal{L}_0+i\omega_m-0^+}}
-\left(\frac{\mathcal{Q}}{\mathcal{L}_0+i\omega_m-0^+}
\mathcal{L}_\mathrm{SB}\right)^3\lim_{\lambda\to0}\lambda^3
\mathcal{R}_m^{(\lambda)}(\tau)\nonumber\\
&=&{-\frac{\mathcal{Q}}{\mathcal{L}_0+i\omega_m-0^+}}\quad(\tau>0),
\label{eqn:KeyFormulaApp}
\end{eqnarray}
which is Eq.\ (\ref{eqn:KeyFormula}) of the text.

A similar argument applies to the case $\tau<0$ to yield
\begin{eqnarray}
\mathcal{R}_m^{(\lambda)}(\tau) &=&\mathcal{Q}
\frac{\e^{(\mathcal{L}_0+i\omega_m)\tau/\lambda^2}-1}%
{\mathcal{L}_0+i\omega_m+0^+}\nonumber\\
&&{}+\lambda\frac{\mathcal{Q}}{\mathcal{L}_0+i\omega_m+0^+}
\e^{i\omega_m\tau/\lambda^2} \rint_0^{\tau/\lambda^2}\d t\,
\e^{\mathcal{L}_0(\tau/\lambda^2-t)}\mathcal{Q}
\mathcal{L}_\mathrm{SB}\mathcal{Q}\e^{\mathcal{L}_0't}\nonumber\\
&&{}-\lambda\frac{\mathcal{Q}}{\mathcal{L}_0+i\omega_m+0^+}
\mathcal{L}_\mathrm{SB}\mathcal{R}_m^{(\lambda)}(\tau)\quad(\tau<0),
\end{eqnarray}
and
\begin{equation}
\lim_{\lambda\to0}\mathcal{R}_m^{(\lambda)}(\tau)
={-\frac{\mathcal{Q}}{\mathcal{L}_0+i\omega_m+0^+}}\quad(\tau<0).
\end{equation}

\subsection{The van Hove Limit of $\mathcal{Q}\e^{\mathcal{L}_0'\tau/\lambda^2}$}
Using the expansion (\ref{eqn:Perturbation}), we have
\begin{equation}
\mathcal{Q}\e^{\mathcal{L}_0'\tau/\lambda^2}
=\mathcal{Q}\e^{\mathcal{L}_0\tau/\lambda^2}
+\lambda\rint_0^{\tau/\lambda^2}\d t\,
\e^{\mathcal{L}_0(\tau/\lambda^2-t)}
\mathcal{Q}\mathcal{L}_\mathrm{SB}
\mathcal{Q}\e^{\mathcal{L}_0't}.
\end{equation}
The convolution in the second term is the same as that discussed in
(\ref{eqn:Convolution}), which is bounded for any $\tau$ and
$\lambda$, provided the projection $\mathcal{P}$ is the
eigenprojection $\Pi_0$, and hence the second term vanishes in van
Hove's limit $\lambda\to0$. Since the right projection
$\mathcal{Q}=1-\Pi_0$ removes the point spectrum of $\mathcal{L}_0$,
the first term disappears as $\tau/\lambda^2\to\infty$ due to
Riemann--Lebesgue's lemma. Therefore,
$\mathcal{Q}\e^{\mathcal{L}_0'\tau/\lambda^2}$ decays in van Hove's
limit, yielding (\ref{eqn:PropQecay}).




\end{document}